\documentclass{sig-alternate}
\usepackage[utf8]{inputenc}

\usepackage{mathptmx} % This is Times font
\usepackage{fancyhdr}
\usepackage[normalem]{ulem}
\usepackage[hyphens]{url}
\usepackage[sort,nocompress]{cite}
\usepackage[final]{microtype}
\usepackage{flushend}

\usepackage{qcircuit}
\usepackage{braket}
\usepackage{xcolor}
\usepackage{float}
\usepackage{caption}
\usepackage{subcaption}
\usepackage{dblfloatfix}

\usepackage{array}
\newcolumntype{P}[1]{>{\raggedright\arraybackslash}p{#1}} % https://tex.stackexchange.com/a/7348

\usepackage[bookmarks=true,breaklinks=true,colorlinks,urlcolor=blue]{hyperref}

\title{Quantum Fan-out: Circuit Optimizations and Technology Modeling}

\author{
Pranav Gokhale\thanks{\href{mailto:pranav@super.tech}{pranav@super.tech}} \\ University of Chicago \and
Samantha Koretsky \\ University of Chicago
\and Shilin Huang \\ Duke University
\and Swarnadeep Majumder \\ Duke University \and Andrew Drucker \\ University of Chicago \and Kenneth R. Brown \\ Duke University \and Frederic T. Chong \\ University of Chicago \\}

\begin{document}
\maketitle
\pagestyle{plain}

%%%%%% -- PAPER CONTENT STARTS-- %%%%%%%%

\begin{abstract}
Instruction scheduling is a key compiler optimization in quantum computing, just as it is for classical computing. Current schedulers optimize for data parallelism by allowing simultaneous execution of instructions, as long as their qubits do not overlap. However, on many quantum hardware platforms, instructions on overlapping qubits can be executed simultaneously through \textit{global interactions}. For example, while fan-out in traditional quantum circuits can only be implemented sequentially when viewed at the logical level, global interactions at the physical level allow fan-out to be achieved in one step. We leverage this simultaneous fan-out primitive to optimize circuit synthesis for NISQ (Noisy Intermediate-Scale Quantum) workloads. In addition, we introduce novel quantum memory architectures based on fan-out.

Our work also addresses hardware implementation of the fan-out primitive. We perform realistic simulations for trapped ion quantum computers. We also demonstrate experimental proof-of-concept of fan-out with superconducting qubits. We perform depth (runtime) and fidelity estimation for NISQ application circuits and quantum memory architectures under realistic noise models. Our simulations indicate promising results with an asymptotic advantage in runtime, as well as 7--24\% reduction in error. 

\end{abstract}

\section{Introduction} \label{sec:introduction}
Instruction scheduling is a powerful compiler technique in both classical and quantum computing. In the classical realm, scheduling techniques such as pipelining, Single Instruction Multiple Data (SIMD), and Out-of-order execution have led to continued gains in processing power. These scheduling techniques are designed to preserve a program's logical correctness by respecting constraints known as \textit{hazards}.

Just as in the classical setting, quantum computing is also amenable to instruction scheduling. In fact, due to the short lifetimes of qubits in the NISQ (Noisy Intermediate-Scale Quantum) era \cite{preskill2018quantum}, scheduling to reduce latency is critical for successful program execution. The potential of quantum instruction scheduling was recently exemplified by Google's Quantum Supremacy result \cite{arute2019quantum}, which experimentally demonstrated a task soluble in seconds on a 53 qubit computer that is argued to likely require days \cite{pednault2019leveraging} on a supercomputer. A core aspect of the Supremacy result is a coupler activation schedule that maximizes simultaneous quantum resource utilization.

A number of papers \cite{guerreschi2018two, guerreschi2019scheduler, metodi2006scheduling, javadiabhari2015scaffcc} in the architecture community have studied quantum scheduling, inspired by techniques from the classical setting. One principle underlying these papers is exclusive activation: a qubit can be involved in at most one operation per timestep \cite{guerreschi2018two}. In architectural terms, this is a \textit{structural hazard} \cite{hennessy2011computer}. Under exclusive activation, schedulers optimize for data parallelism by simultaneously executing instructions on disjoint qubits. However, there are natural limits to such schedulers, since instructions on overlapping qubits must be serialized.

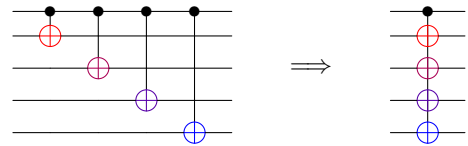
\begin{figure}[b]
$$
\Qcircuit @C=1em @R=0.4em {
& \Ctrl{1} & \Ctrl{2} & \Ctrl{3} & \Ctrl{4} & \qw &&&&&&& \Ctrl{4}    & \qw \\
& \targ{\color{blue!0!red}}    & \qw      & \qw      & \qw   & \qw       &&&&&&& \targ{\color{blue!0!red}}   & \qw \\
& \qw      & \targ{\color{blue!33!red}}    & \qw      & \qw     & \qw     &&&\Longrightarrow&&&& \targ{\color{blue!33!red}}  & \qw \\
& \qw      & \qw      & \targ{\color{blue!66!red}}    & \qw     & \qw    &&&&&&& \targ{\color{blue!66!red}}  & \qw \\
& \qw      & \qw      & \qw      & \targ{\color{blue!100!red}}                       & \qw                   &&&&&&& \targ{\color{blue!100!red}} & \qw 
}
$$
    \caption{Device level fan-out allows a NOT to the bottom four targets iff the top control is on. While exclusive activation induces serialization (left), quantum hardware can implement fan-out simultaneously (right) in a single step.}
    \label{fig:building_block}
\end{figure}

Our work begins with a simple but consequential observation: the structural hazard of exclusive activation is not actually enforced by most quantum hardware. In fact, it can be \textit{more} natural for a quantum processor to simultaneously execute multiple operations on shared qubits through \textit{global interactions}. The building block of our work is the fan-out operation depicted in Figure~\ref{fig:building_block}. This operation can be understood purely classically. The depiction on the left in Figure~\ref{fig:building_block} has four CNOT (Controlled-NOT) gates, each comprising a control ($\bullet$) and a target ($\oplus$). The target is flipped iff the control qubit is 1. This operation performs fan-out for classical input states: when the targets are initialized to 0, the state of the control gets copied to the targets.

While exclusive activation would serialize the four CNOT instructions as depicted on the left, underlying quantum hardware can naturally perform these interactions simultaneously, as depicted on the right. This form of Single Instruction Multiple Data (SIMD) parallelism arises only after discarding structural hazards that don't manifest in hardware. As we demonstrate later, the fan-out building block generalizes to efficiently-scheduled circuit synthesis for the ubiquitous Controlled-$U$ operation. Henceforth in this paper, fan-out will refer to simultaneous operation on the right of Figure~\ref{fig:building_block}.

We begin in Section~\ref{sec:background} with sufficient background on quantum computation so that this paper is self-contained. Section~\ref{sec:prior_work} surveys relevant prior work. The three subsequent sections capture our core contributions:
\begin{itemize}   \setlength\itemsep{-0.25em}
    \item Section~\ref{sec:universality}: We generalize the simultaneous fan-out primitive into a \textbf{circuit synthesis procedure to schedule Controlled-$\mathbf{U}$} operations with an asymptotic depth advantage.
    \item Section~\ref{sec:applications}: We leverage this circuit synthesis procedure to \textbf{optimize NISQ circuits}, and we introduce \textbf{novel quantum memory architectures}.
    \item Section~\ref{sec:trapped_ion}: We perform \textbf{technology modeling} of simultaneous fan-out on trapped ion qubits.
\end{itemize}

Section~\ref{sec:results} presents results for several benchmarks. Section~\ref{sec:superconducting} proposes an implementation of fan-out on superconducting qubits and demonstrates experimental proof-of-concept. Sections~\ref{sec:conclusion} concludes. % To maintain a focus on architectural aspects, we omit details about benchmarks and hardware physics. Our accompanying open source repository (currently anonymized) is well-documented and delves into these aspects. Moreover, to aid understanding, we link to interactive in-browser demos in Quirk \cite{gidney2016quirk} for important circuits.
\section{Background} \label{sec:background}

In order to keep this paper self-contained, we begin with necessary background on quantum computing. To maintain accessibility, we emphasize the circuit model of quantum computing, rather than its linear algebraic formulation.

\subsection{Qubits}
A qubit (quantum bit) is defined by two states, denoted $\ket{0}$ and $\ket{1}$. Just as classical bits can be implemented by a variety of underlying physical representations like magnetization in disks or capacitor charge in RAM, qubits can be fabricated from a variety of underlying quantum technologies. This includes discrete charge levels in superconducting qubits or motional modes in trapped ion qubits. 

The state of a qubit can be written as the linear combination $a\ket{0} + b\ket{1}$, subject to a normalization condition, $|a|^2 + |b|^2 = 1$. This is richer state space that can be captured by a classical bit, which is either $\ket{0}$ or $\ket{1}$. For example, $\frac{1}{\sqrt{2}}[\ket{0} + \ket{1}]$ is a \textit{superposition} qubit state: the qubit has equal components in $\ket{0}$ and $\ket{1}$.

\begin{figure}[h]
     \centering
     \begin{subfigure}[b]{0.06\textwidth}
         \centering
$$\Qcircuit @C=0.3em @R=1em { & \Targ & \qw }$$
\caption{NOT} \label{subfig:not} \end{subfigure}
\hfill
\begin{subfigure}[b]{0.09\textwidth} \centering $$\Qcircuit @C=0.5em @R=1em { & \gate{H} & \qw \\ } $$
\caption{Hadamard} \label{subfig:h} \end{subfigure}
\hfill
\begin{subfigure}[b]{0.1\textwidth}
         \centering $$\Qcircuit @C=0.5em @R=1em {& \gate{R_z(\theta)} & \qw \\ }$$
\caption{$R_z$ rotation} \label{subfig:rz} \end{subfigure}
\hfill
\begin{subfigure}[b]{0.08\textwidth}
         \centering $$\Qcircuit @C=0.5em @R=0.65em { & \Ctrl{1} & \qw \\ & \Targ    & \qw } $$
\caption{CNOT} \label{subfig:cnot} \end{subfigure}
\hfill
\begin{subfigure}[b]{0.12\textwidth}
         \centering $$\Qcircuit @C=0.5em @R=0.3em { & \Ctrl{1} & \qw \\ & \gate{U}    & \qw } $$
\caption{Controlled-$U$} \label{subfig:c-u} \end{subfigure}
        \caption{Gates used in this paper.} \label{fig:gates}
\end{figure}
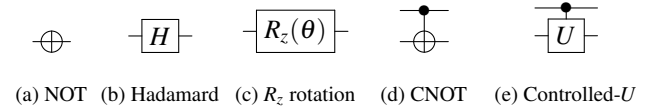

The state of a qubit can be changed by a gate. Figure~\ref{fig:gates} depicts the gates we use in this paper. Each gate has an input wire(s) and an output wire(s). The first gate is just the classical NOT gate, which interchanges $\ket{0}$ and $\ket{1}$. The next gate is the Hadamard gate, which is an intrinsically quantum gate that creates superposition. For example, applying $H$ to a $\ket{0}$ creates the equal superposition $\frac{1}{\sqrt{2}}[\ket{0} + \ket{1}]$. The $R_z(\theta)$ gate is another quantum gate which applies a phase. Phase is helpfully visualized as a $\theta$ displacement in longitude on a sphere; however, for our purposes its underlying meaning is unimportant. Gates (d)~and~(e) act on pairs of qubits (wires). The CNOT is a Controlled-NOT, which applies a NOT to the bottom qubit, iff the top qubit is $\ket{1}$. The top qubit itself is unaffected. The Controlled-$U$ is simply a generalization, where $U$ represents some single qubit gate that is only activated if the top qubit is $\ket{1}$.

While a qubit can carry a rich state space, it snaps to either $\ket{0}$ or $\ket{1}$ upon measurement. This process is fundamentally stochastic: the probability of measuring $\ket{0}$ is $|a|^2$ and $\ket{1}$ is $|b|^2$, which justifies the normalization condition. For example, under the $\frac{1}{\sqrt{2}}[\ket{0} + \ket{1}]$ state, the probability of measuring $\ket{0}$ is $(\frac{1}{\sqrt{2}})^2 = \frac{1}{2}$ which is indeed an equal superposition. The measurement operation is visually denoted as $\Qcircuit @C=0.5em @R=1em {& \meter }$, which terminates a wire.

\subsection{Quantum circuits}
Quantum programs are expressed as quantum circuits which, like Boolean circuits, carry wires representing qubits through a sequence of gates. An example quantum circuit is shown below in Figure~\ref{fig:example_circuit}. It can be read as a timeline from left to right. As indicated, the \textit{width} is the number of qubits the circuit acts on. In addition to data qubits that encode input/output, quantum circuits often use extra \textit{ancilla} qubits that store temporary results. The \textit{depth} is the length of the critical path. Thus, width and depth respectively capture the space and time costs of a quantum circuit.

\begin{figure}[h]
    \centering
$$
\Qcircuit @C=1.7em @R=0.35em {
& && \gate{R_z(\theta_1)}                                    & \Ctrl{1} & \Targ & \qw \\
& \hspace{-4.2em} \raisebox{-1.5em}{\text{width}} && \Targ  & \Targ     & \gate{U} & \qw \\
& && \Targ                                 & \Targ       & \Ctrl{-1}  & \qw \\
& && \gate{R_z(\theta_2)}  & \Ctrl{-1} & \gate{R_z(\theta_3)} & \qw \gategroup{1}{2}{4}{3}{1em}{\{} \gategroup{4}{4}{4}{6}{1em}{_\}} \\
&&&& \raisebox{-2.2em}{\text{depth}}  &            \\
}
$$
    \caption{An example quantum circuit with a width of 4 qubits and a depth of 3 layers.}
    \label{fig:example_circuit}
\end{figure}
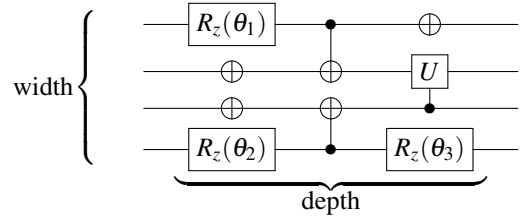

Figure~\ref{fig:example_circuit} exemplifies data parallelism---no qubit is ever idle. Such speedups are especially important in quantum computing because qubits generally have short coherence windows for useful computation.

\subsection{Commutativity}
Every quantum circuit has an underlying program dependency graph that enforces the execution order of gates. Naively, one can construct a program dependency graph that simply adds forward dependencies from each quantum gate to subsequent quantum gates in the circuit-timeline. However, this dependency graph can often be relaxed due to commutativity, where two quantum gates can be applied in either order.

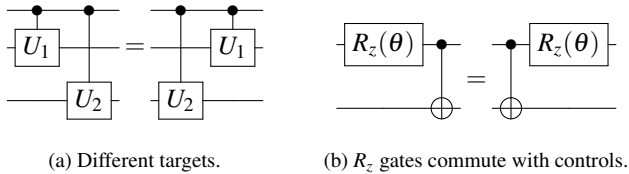
\begin{figure}[h]
     \centering
     \begin{subfigure}[b]{0.2\textwidth}
         \centering
$$\Qcircuit @C=0.3em @R=0.6em {
& \Ctrl{1} & \Ctrl{2} & \qw  && &&& \Ctrl{2} & \Ctrl{1} & \qw \\
& \gate{U_1}    & \qw      & \qw  &&=&&& \qw      & \gate{U_1}    & \qw \\
& \qw      & \gate{U_2}    & \qw  && &&& \gate{U_2}    & \qw      & \qw \\
}
$$
\caption{Different targets.} \label{subfig:different_targets}
     \end{subfigure}
\hfill
     \begin{subfigure}[b]{0.24\textwidth}
         \centering
$$\Qcircuit @C=0.3em @R=0.6em {
& \gate{R_z(\theta)} & \Ctrl{2} & \qw &&&&& \Ctrl{2} & \gate{R_z(\theta)} & \qw \\
&                    &          &     && = \\
& \qw                & \Targ    & \qw &&&&& \Targ    & \qw & \qw \\
}
$$
\caption{$R_z$ gates commute with controls.} \label{subfig:rz_commutativity}
     \end{subfigure}
        \caption{Two commutativity rules encountered in this paper.}
        \label{fig:commutativity}
\end{figure}

Many commutativity relationships exist between gates. For our work, we only rely on the two relationships depicted in Figure~\ref{fig:commutativity}. The left equivalence shows that two Controlled-$U_i$ gates commute when they have different targets. This is clear because controlled gates leave the control qubit unchanged, so their order is unimportant. The right equivalence shows that $R_z$-type gates commute with controls of controlled gates (such as CNOTs). This relationship has no classical analogue, but the underlying intuition is that $R_z$ gates don't affect the $\ket{0}$ vs. $\ket{1}$ balance of a qubit, so their order relative to a control is unimportant. Both of these commutativity rules are used in our Controlled-$U$ circuit synthesis procedure in Section~\ref{sec:universality}.

% \subsection{No-Cloning Theorem}
% A unique and challenging aspect of quantum computing is that in general, a qubit cannot be copied. This restriction is termed the No-Cloning Theorem \cite{wootters1982single}. At a glance, the No-Cloning Theorem appears to conflict our reliance on a Fan-Out operation. However, fan-out of quantum states is distinct from Cloning. For example, fanning out the superposition state $\frac{\ket{0} + \ket{1}}{\sqrt{2}}$ to two target qubits results in the output state
% $$\frac{\ket{000} + \ket{111}}{\sqrt{2}}.$$
% This is different from cloning the state to two targets, which would result in
% \small $$\frac{\ket{000} + \ket{001} + \ket{010} + \ket{011} + \ket{100} + \ket{101} + \ket{110} + \ket{111}}{\sqrt{8}}.$$ \normalsize
\section{Prior Work} \label{sec:prior_work}
Our work builds on top of prior work from the (1) computer architecture, (2) computer science theory, and (3) physics communities. At a high level, the priorities of the work in each community can be characterized as follows:
\begin{enumerate}  \setlength\itemsep{-0.25em}
    \item architects have devised intelligent schedulers/circuit synthesis tools, but they assume a false structural hazard by overlooking global interactions
    \item theorists have devised intelligent circuit constructions assuming global gates, but they don't consider NISQ workloads or device-level operation
    \item physicists have studied global interactions, but usually in an ad hoc fashion separated from computation and NISQ workloads
\end{enumerate}

Our work unites insights from all three disciplines to devise a circuit synthesis tool that leverages global interactions to accelerate NISQ workloads.

\subsection{Computer Architecture}

Amongst architects, a number of papers \cite{guerreschi2018two, guerreschi2019scheduler, metodi2006scheduling, javadiabhari2015scaffcc, venturelli2018compiling, heckey2015compiler} have studied instruction scheduling in quantum computers. These papers all assume some structural hazard against simultaneous execution of overlapping qubits. \cite{guerreschi2018two, guerreschi2019scheduler} provides the most formal description of this hazard, terming it as the principle of exclusive activation which forbids a qubit from being involved in more than one operation per timestep.
Moreover, hardware-dependent considerations such as crosstalk \cite{li2019towards, murali2020software} further narrow the scope of when operations can be parallelized. For example, on superconducting hardware, \texttt{CNOT(a,b); CNOT(c,d);} may be forbidden simultaneously if they are neighboring pairs, even though the \texttt{CNOT} gates are disjoint.

In other architectural work such as \cite{metodi2006scheduling} and \cite{heckey2015compiler}, the authors provide examples for obtaining data parallelism on disjoint instructions. However, in both papers, the examples ultimately incur serialization upon encountering gates on overlapping qubits. As we will demonstrate in Section~\ref{sec:universality}, this serialization is unnecessary.

Finally, \cite{javadiabhari2015scaffcc} describes exclusive activation as a data dependency, since the no-cloning theorem \cite{wootters1982single} prevents copying a qubit to participate in multiple instructions simultaneously. This is indeed a valid perspective. Regardless, we will demonstrate that the underlying problem is in fact addressable with the fan-out primitive.

% Before proceeding, we also note that the structural hazard against exclusive activation is quite familiar in classical architecture. For instance, classical memory is typically treated as single-port or finitely multi-port, but not infinitely multi-port. The unbounded multi-port nature of global interactions is quite a surprising opportunity in this regard.

\subsection{Computer Science Theory}
Quantum fan-out has also been studied from a complexity theory lens. \cite{hoyer2005quantum} proved that the QNC\textsuperscript{0}\textsubscript{f} class of circuits with unbounded fan-out is powerful for fault-tolerant applications such as Shor's Algorithm \cite{shor1999polynomial} for factoring. Other applications of fan-out to arithmetic operations such as addition, OR, and modulus are considered in \cite{takahashi2009quantum}, \cite{takahashi2016collapse}, and \cite{green2001counting} respectively. Finally, \cite{takahashi2014hardness} shows that under widely-held complexity theory assumptions, fan-out in quantum circuits can increase the hardness of classical simulability. Our work revisits these theory results with NISQ workloads and underlying device physics in mind.

\subsection{Physics} \label{subsec:prior_physics}
The engineering of global interactions on $N$ qubits has been well studied in device physics communities. A common benchmark for global interactions is the preparation of the \textit{GHZ state} \cite{greenberger1989going}, a task which is essentially equivalent to fan-out. Experimentally, global interactions have been used to prepare the GHZ state on a variety of leading qubit technologies including Trapped Ion \cite{lu2019global}, Neutral Atom \cite{omran2019generation}, and NMR \cite{dogra2015experimental}. Implementation on NV center qubits has been proposed as well \cite{goldstein2011environment}. Notably, superconducting qubits, which are the current leader in hardware scale, were not previously known to support simultaneous overlapping interactions. However, in Section~\ref{sec:superconducting}, we experimentally demonstrate simultaneous fan-out on superconducting qubits.

Global interactions have already been noted by physicists for their application to Hamiltonian simulation, an important quantum algorithm. As early as 2005 \cite{zeng2005measuring}, it was noted that global interactions enable constant depth parity measurement, an important building block for Hamiltonian simulation. Later work \cite{leibfried2018efficient, maslov2018use} further optimized and clarified the procedure. Recently this year, three papers \cite{groenland2020sequences, rasmussen2020single, yu2020scalability} have applied global interactions to building blocks of longer-term fault tolerant quantum computers. These papers demonstrate that the Generalized Toffoli operation can be performed in constant time with global interactions, whereas otherwise linear or logarithmic depth is required \cite{gokhale2019asymptotic, gokhale2020extending}.

Very recent papers have adopted an interdisciplinary approach, combining insights from physics and architecture. For example, \cite{maslov2018use}---which inspired our work---describes fan-out as SIMD parallelism. Also, a recent trapped ion hardware paper \cite{grzesiak2019efficient} describes global interactions as a form of Multiple Instruction Multiple Data (MIMD) parallelism. Our work continues this architectural perspective, while also focusing on NISQ circuit optimizations and further refining the underlying technology models based on recent experimental developments.
\section{Controlled-U Synthesis} \label{sec:universality}

The basic building block of our work is the fan-out operation depicted on the right side of Figure~\ref{fig:building_block}. Two important considerations arise in evaluating the applicability of fan-out. The first is whether the simultaneous implementation via global interactions truly achieves a linear speedup over serialized CNOTs\footnote{Fan-out can also be implemented in logarithmic depth via a recursive tree structure, but we focus on the serialized case which is dominant in prior work.}. As described in Sections~\ref{sec:trapped_ion} and \ref{sec:superconducting}, experimental results from hardware assert this is indeed the case. The second consideration is how fidelity is affected by simultaneous fan-out versus serialized CNOTs. Our results in Sections~\ref{sec:trapped_ion} and \ref{sec:superconducting} indicate a modest improvement in fidelity from simultaneous fan-out.

We focus on a circuit synthesis procedure that uses fan-out to optimize the Controlled-$U$ operation, described below. This operation is ubiquitous in NISQ algorithms, and each application in Section \ref{sec:applications} is an instance of Controlled-$U$. As we will describe in this section, our circuit synthesis procedure yields a Controlled-$U$ implementation that is scheduled to align CNOT gates into a single fan-out step. This yields asymptotic improvements in circuit depth. Our code was implemented as a fork of Qiskit Terra \cite{Qiskit}.

\begin{figure}[h]
    \centering
    \includegraphics[width=0.48\textwidth]{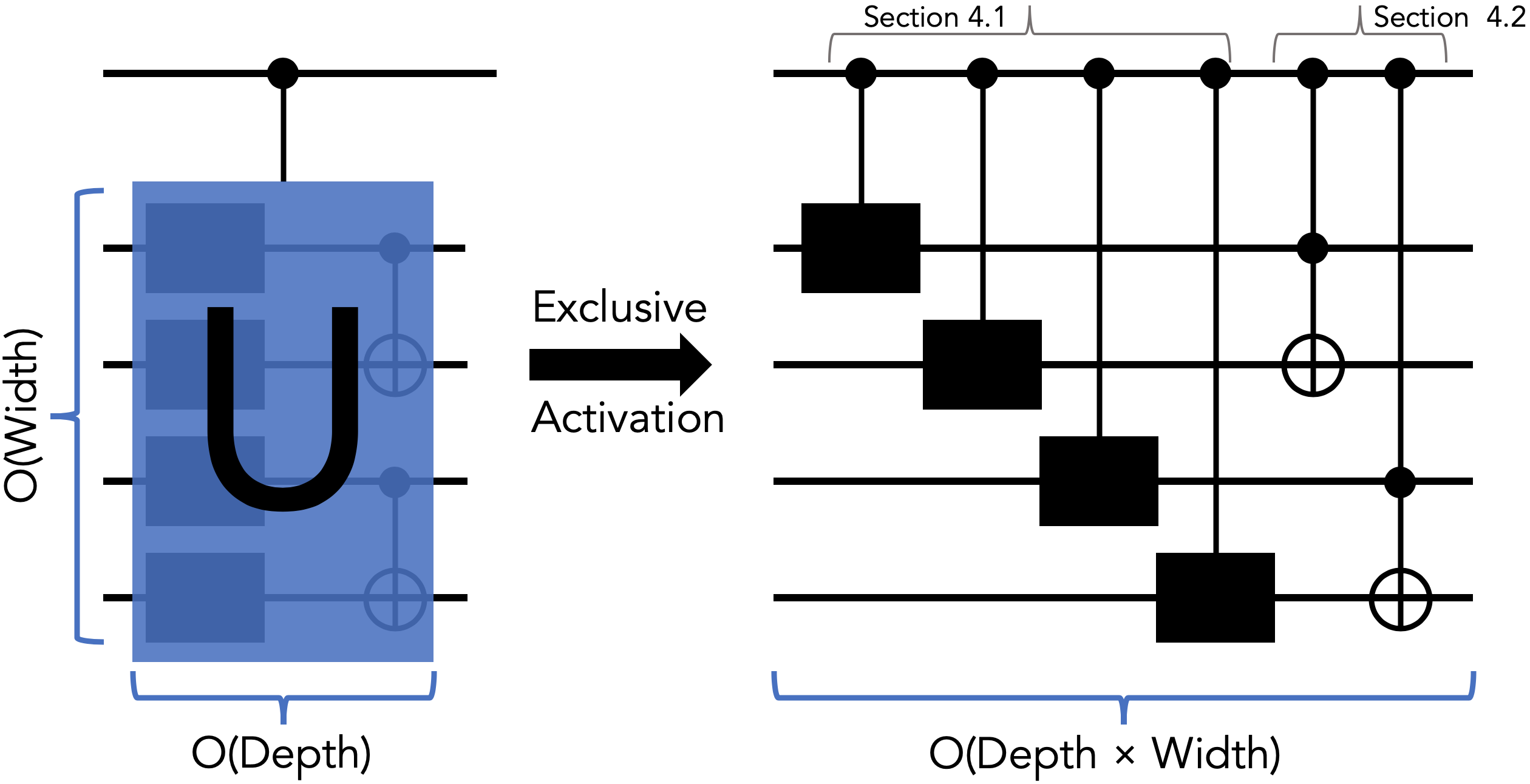}
    \caption{Left: general form of controlled-$U$. Right: under exclusive activation, adding the control induces serialization and multiplies the effective Depth by the Width.} \label{fig:Universality}
\end{figure}

The controlled-$U$ operation is depicted at the left of Figure~\ref{fig:Universality}. As in other controlled operations like CNOT, the $U$ operation should be applied if and only if the top control qubit is $\ket{1}$. However, unlike the single-qubit $U$ in Figure~\ref{subfig:c-u}, here we consider the case when $U$ is an operation on multiple qubits. Therefore, $U$ itself has a decomposition into gates, shown under the blue overlay. Our results are applicable for any decomposition basis, but we focus on the decomposition into the universal set of single-qubit + CNOT gates, since quantum algorithms are typically expressed in this form. In the example, $U$ has a width of four qubits and a depth of two layers. The first layer contains four disjoint single qubit gates, and the second layer contains two disjoint CNOTs.

Under exclusive activation, implementation of Controlled-$U$ is bottlenecked by the dependence of each controlled gate on the single control qubit. Thus, the parallel two-layer implementation of $U$ collapses into a serial implementation of Controlled-$U$ as depicted at the right of Figure~\ref{fig:Universality}. The amount of serialization is proportional to the width of $U$, so that the effective depth of a Controlled-$U$ operation under serialization is $O(\text{Depth } \times \text{ Width})$. In many workloads, the width greatly exceeds depth, so this serialization is very harmful.

It is not immediately obvious how fan-out can help speed up Controlled-$U$. Whereas fan-out is a SIMD operation, Controlled-$U$ is a MIMD operation, since the gates in $U$ are arbitrary. However, we can resolve this difficulty be decomposing gates into a form amenable to `alignment' of CNOTs into a single fan-out step. This circuit synthesis procedure has two underlying cases. The first, Shared-Control Single Qubit Gates, supports the simultaneous execution of multiple Controlled-$U_i$ gates with a shared control qubit. This procedure applies to the first layer of $U$ in Figure~\ref{fig:Universality}. The second, Shared-Control Toffoli's, supports the simultaneous execution of multiple Controlled-CNOTs with a shared control qubit. These double-controlled NOTs are referred to as Toffoli's. The Shared-Control Toffoli's case applies to the second layer of $U$ in Figure~\ref{fig:Universality}.

In practice, arbitrary $U$'s will also contain mixed layers that contain both single-qubit gates and CNOTs. This general case can be handled by unifying the synthesis procedures for Shared-Control Single Qubit Gates and Shared-Control Toffoli's. It is not presented here for brevity, but is implemented in our code.

Table~\ref{tab:controlled_U_cost} compares the time (depth) and space (ancilla qubits) costs of implementing Controlled-$U$. Our work, which uses fan-out, is optimal with $O(D)$ depth (and very small constants) and 0 ancilla qubits. The status quo approach of serialization incurs $O(ND)$ depth which is harmful because $N >> D$ in many applications. Past work in \cite{hoyer2005quantum} and \cite{martinez2016compiling} has proposed alternative approaches for parallelizing circuits using global interactions. In the best case, where a ``basis-change'' is cheap and efficiently computable, \cite{hoyer2005quantum} matches our $O(D)$ depth. However, it is extremely expensive in space, requiring $O(N^2)$ ancilla qubits. %Again, this is impractical since the relevant $N$ typically spans the entire device.
Finally, \cite{martinez2016compiling} provides a numerical optimization technique for compiling Controlled-$U$ down to the minimal possible depth. In this sense, it could achieve the $O(D)$ lower bound. However, the numerical optimization for compilation has exponential cost---simply defining the optimization problem involves specifying a $2^{N}\times 2^{N}$ sized matrix. Moreover, the optimization itself is expensive, and convergence to $O(D)$ depth is not guaranteed.

\begin{table}[h]
\renewcommand{\arraystretch}{1.1} \centering
\begin{tabular}{l|cc}
         & Depth                & Ancilla Qubits          \\ \hline
\textbf{Our Work (with fan-out)} & $\mathbf{O(D)}$               & $\mathbf{0}$  \\
Serialization & $O(ND)$         & 0  \\
\cite{hoyer2005quantum} (if cheap basis-change) & $O(D)$         & $O(N^2)$  \\
\cite{martinez2016compiling} ($\Omega(2^N)$ compile time) & $O(D)$? & 0 \\
\end{tabular}
\caption{Cost of implementing a controlled-$U$ operation in time (depth) and space (ancilla qubits). $U$ has a depth of $D$ and width of $N$ qubits.}
\label{tab:controlled_U_cost}
\end{table}

While our procedure achieves the best possible asymptotic spacetime costs, it is not as general as \cite{hoyer2005quantum, martinez2016compiling}. Our procedure only addresses the special case of Controlled-$U$ parallelization, whereas \cite{hoyer2005quantum} and \cite{martinez2016compiling} apply to the parallelization of any commuting gates or the depth reduction of any unitary, respectively. Nonetheless, our specialization is justified because the Controlled-$U$ template is ubiquitous in NISQ workloads.

\subsection{Shared-Control Single Qubit Gates}
Here, we consider how to simultaneously execute controlled single-qubit gates with a shared control, as in the first layer of $U$ in Figure~\ref{fig:Universality}. This is a form of MIMD parallelism with overlapping data, but we only have access to the fan-out SIMD primitive. However, we can make progress by invoking the following well-known identity \cite{nielsen2002quantum} for decomposing controlled single-qubit gates. It shows that for any single-qubit gate $U$, the Controlled-$U$ operation can be implemented by using CNOT as the only two-qubit gate. Specifically there exist (trivially computable) single-qubit gates $A$, $B$, $C$, and an angle $\theta$, such that
$$
\Qcircuit @C=1em @R=0.6em {
& \Ctrl{1} & \qw &&\raisebox{-2.5em}{=}&&& \qw      & \Ctrl{1} & \qw      & \Ctrl{1} & \gate{R_z(\theta)} & \qw \\
& \gate{U}                & \qw &&&&& \gate{C} & \Targ    & \gate{B} & \Targ & \gate{A}      & \qw
}
$$

Let us consider applying this identity to a small example: attempting to parallelize Controlled-$U_1$ and Controlled-$U_2$ targeting two different qubits. The result is shown below, with colors used for disambiguation.

$$
\Qcircuit @C=0.5em @R=0.6em {
& \Ctrl{1}               & \qw &&&&& \qw      & \Ctrl{1} & \qw      & \Ctrl{1} & \gate{R_z({\color{red} \theta_1})} & \Ctrl{2} & \qw & \Ctrl{2} & \gate{R_z({\color{blue} \theta_2})} & \qw \\
& \gate{\color{blue!0!red} U_1} \qwx[1] & \qw &&=&&& \gate{\color{red} C_1} & \targ{\color{red}}    & \gate{\color{red} B_1} & \targ{\color{red}} & \gate{\color{red} A_1} & \qw & \qw & \qw & \qw & \qw \\
& \gate{\color{blue!100!red} U_2}        & \qw &&&&& \gate{\color{blue} C_2} & \qw & \qw & \qw & \qw & \targ{\color{blue}} & \gate{\color{blue} B_2} & \targ{\color{blue}} & \gate{\color{blue} A_2} & \qw
}
$$

\begin{figure}[h]
$$
\Qcircuit @C=0.5em @R=0.6em {
& \Ctrl{1} & \qw &&&&& \qw                             & \qw & \Ctrl{2}     & \qw & \qw                             & \qw & \Ctrl{1}     & \qw & \gate{R_z({\color{blue} \theta_1} + {\color{red} \theta_2})}      & \qw                       \\
& \gate{\color{blue} U_1} \qwx[1]  & \qw &&=&&& \gate{\color{blue} C_1}   & \qw & \targ{\color{blue}}   & \qw & \gate{\color{blue} B_1}   & \qw & \targ{\color{blue}} \qwx[1]   & \qw & \gate{\color{blue} A_1} & \qw   \\
& \gate{\color{red} U_2} & \qw &&&&& \gate{\color{red} C_2}  & \qw & \targ{\color{red}}  & \qw & \gate{\color{red} B_2}  & \qw & \targ{\color{red}}  & \qw & \gate{\color{red} A_2} & \qw \\
}
$$
    \caption{Simultaneous execution of shared-control single qubit gates, using the fan-out primitive. This decomposition has constant (5 layer) depth, independent of width.}
    \label{fig:controlled_single_qubit_mimd}
\end{figure}
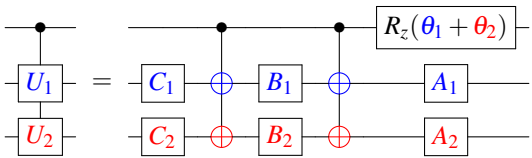

It appears that applying the circuit identity led to minimal improvements---only $C_2$ can slide left to execute simultaneously with controlled-$U_1$. The rest of the blue gates are unable to parallelize, because they are blocked by an apparent dependency on the $R_z(\theta_1)$ gate. However, recalling the commutativity rule in Figure~\ref{subfig:rz_commutativity}, we see that the apparent dependence of the blue CNOTs on the $R_z(\theta_1)$ is actually a false dependence. By commuting the $R_z(\theta_1)$ gate to the end of the circuit, we attain the final result in Figure~\ref{fig:controlled_single_qubit_mimd}.

We have now demonstrated simultaneous execution of shared-control $U_1$ and $U_2$ on overlapping data (top+middle and top+bottom qubits respectively), using the fan-out primitive. This pattern extends \textit{ad infinitum} to more qubits---the total depth will always consist of five layers: two fan-out layers and three single-qubit gate layers. For certain gates, the cost could be reduced even further. For instance, for $U = Z$, it is known that the Controlled-$Z$ operation can be implemented with just a single CNOT \cite{nielsen2002quantum}.
% For brevity, we only present the general case here which already has small constants.

\subsection{Shared-Control Toffoli's}
The second piece needed for optimized Controlled-$U$ synthesis is simultaneous execution of shared-control Toffoli's. Here, we seek to simultaneously execute multiple Toffoli (Controlled-CNOT) gates, where the CNOTs are disjoint but the additional control is shared across the CNOTs, as in the second layer of $U$ in Figure~\ref{fig:Universality}. Since Toffoli is a three-qubit operation, it must first be decomposed into single-qubit gates and CNOTs. The standard \cite{nielsen2002quantum} decomposition is shown next. $T$ and $T^{\dagger}$ are shorthand for $R_z(\frac{\pi}{8})$ and $R_z(\frac{-\pi}{8})$ respectively.

$$
\Qcircuit @C=0.4em @R=0.6em {
& \Ctrl{2} &\qw  &&&&& \qw      & \qw      & \qw            & \Ctrl{2} & \qw      & \qw      & \qw           & \Ctrl{2} & \qw      & \Ctrl{1} & \gate{T} \gategroup{1}{18}{2}{18}{.7em}{--}      & \Ctrl{1} & \qw \\
& \Ctrl{1} &\qw &&=&&& \qw      & \Ctrl{1} & \qw            & \qw      & \qw      & \Ctrl{1} & \qw           & \qw      & \gate{T} & \Targ    & \gate{T^{\dagger}} & \Targ & \qw \\
& \Targ    &\qw  &&&&& \gate{H} & \Targ    & \gate{T^{\dagger}} & \Targ     & \gate{T} & \Targ    & \gate{T^{\dagger}} & \Targ    & \gate{T} & \gate{H} & \qw           & \qw & \qw \\
}
$$

The boxed group with $T$ and $T^{\dagger}$ is one example of data parallelism. This level of data parallelism is referred to as a coarse-grained schedule in past architectural work \cite{heckey2015compiler}. Next, let us consider applying the Toffoli decomposition to a small example: attempting to simultaneously execute two shared-control Toffoli's, where the CNOTs are disjoint. This exact example is also considered in Figure 4 of \cite{heckey2015compiler}. The result is shown below, with colors used again for disambiguation.

$$ \scriptsize
\Qcircuit @C=0.3em @R=0.8em {
& \Ctrl{4} &\qw  &&&&& \qw      & \qw      & \qw            & \Ctrl{2} & \qw      & \qw      & \qw           & \Ctrl{2} & \qw      & \Ctrl{1} & \gate{\color{red} T}      & \Ctrl{1} & \Ctrl{4} & \qw      & \qw      & \qw           & \Ctrl{4} & \qw      & \Ctrl{3} & \gate{\color{blue} T}      & \Ctrl{3} & \qw \\
& \ctrl{1}{\color{red}} &\qw  &&&&& \qw      & \ctrl{1}{\color{red}} & \qw            & \qw      & \qw      & \ctrl{1}{\color{red}} & \qw           & \qw      & \gate{\color{red} T} & \targ{\color{red}}    & \gate{\color{red} T^{\dagger}} & \targ{\color{red}} & \qw & \qw & \qw & \qw & \qw & \qw & \qw & \qw & \qw & \qw \\
& \targ{\color{red}}    &\qw &&=&&& \gate{\color{red} H} \gategroup{2}{8}{5}{10}{.7em}{--} & \targ{\color{red}}    & \gate{\color{red} T^{\dagger}}  &\targ{\color{red}}     & \gate{\color{red} T} & \targ{\color{red}}    & \gate{\color{red} T^{\dagger}} & \targ{\color{red}}    & \gate{\color{red} T} & \gate{\color{red} H} & \qw           & \qw & \qw & \qw & \qw & \qw & \qw & \qw & \qw & \qw & \qw & \qw \\
& \ctrl{1}{\color{blue}} &\qw  &&&&& \qw      & \ctrl{1}{\color{blue}} & \qw            & \qw      & \qw      & \qw      & \qw           & \qw      & \qw      & \qw      & \qw           & \qw      & \qw      & \qw      & \ctrl{1}{\color{blue}} & \qw           & \qw      & \gate{\color{blue} T} & \targ{\color{blue}}    & \gate{\color{blue} T^{\dagger}} & \targ{\color{blue}} & \qw \\
& \targ{\color{blue}}    &\qw  &&&&& \gate{\color{blue} H} & \targ{\color{blue}}    & \gate{\color{blue} T^{\dagger}}  & \qw      & \qw      & \qw      & \qw           & \qw      & \qw      & \qw      & \qw           & \qw      &\targ{\color{blue}}     & \gate{\color{blue} T} & \targ{\color{blue}}    & \gate{\color{blue} T^{\dagger}} & \targ{\color{blue}}    & \gate{\color{blue} T} & \gate{\color{blue} H} & \qw           & \qw & \qw \\
}
$$

As indicated by the boxed layers, only three gates from the blue Toffoli were able to parallelize with the execution of the red Toffoli. This level of parallelization, which results in 21 layers of depth, is referred to as fine-grained scheduling in \cite{heckey2015compiler}. While it is slightly better than coarse-grained scheduling, it still linearly serializes the depth. However, we can again leverage commutativity relationships to proceed further and exploit our fan-out primitive.

Notice that the dependency between the right-most red CNOT and the subsequent blue CNOT is in fact a false dependency. These two gates commute per the rule in Figure~\ref{subfig:different_targets} since their targets are different. After transposing the two gates, we encounter a $T$ gate that commutes with the control of the blue CNOT, per the rule in Figure~\ref{subfig:rz_commutativity}. Repeating such commutative transpositions, we can push the blue CNOT to the left to align into a single fan-out. The rest of the blue circuit can be handled similarly, resulting in the final form presented in Figure~\ref{fig:toffoli_parallelization}. Since $T = R_z(\frac{\pi}{8})$, the $T \times T$ gate at the top right is just a single $R_z(\frac{\pi}{4})$ gate.

\begin{figure}[h]
$$ \footnotesize
\Qcircuit @C=0.4em @R=0.6em {
& \Ctrl{4} &\qw  &&&&&&& \qw      & \qw      & \qw            & \Ctrl{4} & \qw      & \qw      & \qw           & \Ctrl{4} & \qw      & \Ctrl{3} & \gate{{\color{red} T} \times {\color{blue} T}}      & \Ctrl{3} & \qw \\
& \ctrl{1}{\color{red}} &\qw  &&&&&&& \qw      & \ctrl{1}{\color{red}} & \qw            & \qw      & \qw      & \ctrl{1}{\color{red}} & \qw           & \qw      & \gate{\color{red} T} & \targ{\color{red}}    & \gate{\color{red} T^{\dagger}} & \targ{\color{red}} & \qw \\
& \targ{\color{red}}    &\qw &&&=&&&& \gate{\color{red} H} & \targ{\color{red}}    & \gate{\color{red} T^{\dagger}}  &\targ{\color{red}}     & \gate{\color{red} T} & \targ{\color{red}}    & \gate{\color{red} T^{\dagger}} & \targ{\color{red}}    & \gate{\color{red} T} & \qw & \gate{\color{red} H} & \qw  & \qw \\
& \ctrl{1}{\color{blue}} &\qw  &&&&&&& \qw      & \ctrl{1}{\color{blue}} & \qw            & \qw  & \qw & \ctrl{1}{\color{blue}}  & \qw      & \qw     & \gate{\color{blue} T} & \targ{\color{blue}}    & \gate{\color{blue} T^{\dagger}} & \targ{\color{blue}} & \qw \\
& \targ{\color{blue}}    &\qw  &&&&&&& \gate{\color{blue} H} & \targ{\color{blue}}    & \gate{\color{blue} T^{\dagger}}  & \targ{\color{blue}} & \gate{\color{blue} T}  & \targ{\color{blue}} & \gate{\color{blue} T^{\dagger}} & \targ{\color{blue}}    & \gate{\color{blue} T} & \qw & \gate{\color{blue} H}  & \qw & \qw \\
}
$$
\caption{Simultaneous execution of shared-control Toffoli's using the fan-out primitive. This decomposition has constant (12 layer) depth, independent of width. \href{https://algassert.com/quirk\#circuit=\%7B\%22cols\%22\%3A\%5B\%5B\%22Counting5\%22\%5D\%2C\%5B1\%2C1\%2C\%22H\%22\%2C1\%2C\%22H\%22\%5D\%2C\%5B1\%2C\%22\%E2\%80\%A2\%22\%2C\%22X\%22\%5D\%2C\%5B1\%2C1\%2C1\%2C\%22\%E2\%80\%A2\%22\%2C\%22X\%22\%5D\%2C\%5B1\%2C1\%2C\%22Z\%5E-\%C2\%BC\%22\%2C1\%2C\%22Z\%5E-\%C2\%BC\%22\%5D\%2C\%5B\%22\%E2\%80\%A2\%22\%2C1\%2C\%22X\%22\%2C1\%2C\%22X\%22\%5D\%2C\%5B1\%2C1\%2C\%22Z\%5E\%C2\%BC\%22\%2C1\%2C\%22Z\%5E\%C2\%BC\%22\%5D\%2C\%5B1\%2C\%22\%E2\%80\%A2\%22\%2C\%22X\%22\%5D\%2C\%5B1\%2C1\%2C1\%2C\%22\%E2\%80\%A2\%22\%2C\%22X\%22\%5D\%2C\%5B1\%2C1\%2C\%22Z\%5E-\%C2\%BC\%22\%2C1\%2C\%22Z\%5E-\%C2\%BC\%22\%5D\%2C\%5B\%22\%E2\%80\%A2\%22\%2C1\%2C\%22X\%22\%2C1\%2C\%22X\%22\%5D\%2C\%5B1\%2C\%22Z\%5E\%C2\%BC\%22\%2C\%22Z\%5E\%C2\%BC\%22\%2C\%22Z\%5E\%C2\%BC\%22\%2C\%22Z\%5E\%C2\%BC\%22\%5D\%2C\%5B\%22\%E2\%80\%A2\%22\%2C\%22X\%22\%2C1\%2C\%22X\%22\%5D\%2C\%5B\%22Z\%5E\%C2\%BD\%22\%2C\%22Z\%5E-\%C2\%BC\%22\%2C\%22H\%22\%2C\%22Z\%5E-\%C2\%BC\%22\%2C\%22H\%22\%5D\%2C\%5B\%22\%E2\%80\%A2\%22\%2C\%22X\%22\%2C1\%2C\%22X\%22\%5D\%5D\%7D}{Quirk Link}.}
\label{fig:toffoli_parallelization}
\end{figure}

The design in Figure~\ref{fig:toffoli_parallelization} extends naturally to more qubits. Regardless of the number of qubits, the depth of the circuit is always 12 layers. Since the depth of a single Toffoli operation is also 12 layers, this means that our shared-control Toffoli's synthesis is optimal. For the circuits we will encounter in the following sections, the number of Toffoli's spans the entire circuit. Therefore the depth cost of the other approaches is $O(N)$, versus our $12 = O(1)$ constant depth.

The combination of simultaneous shared-control single qubit gates and Toffoli's enables  a depth-optimized execution schedule for any Controlled-$U$. Moreover, the multiplicative constants for our circuit synthesis are small. Shared-control single qubit gates incur a depth of just 5 layers, which matches worst case depth. Shared-control Toffoli's incur no depth expansion relative to a single Toffoli and are thus optimal. The resulting Controlled-$U$ circuit synthesis procedure is implemented in our code. In the following section, we apply the Controlled-$U$ procedure to optimize several NISQ-important quantum circuits, which are all fundamentally Controlled-$U$ operations. While our approach is already asymptotically optimal with low constants, in some cases we can reduce the depth constants even further. This is exemplified by the SWAP Test, which we discuss next.
\section{Applications} \label{sec:applications}
We now examine how Controlled-$U$ circuit synthesis can be leveraged to optimize NISQ circuits. We also apply fan-out to develop novel quantum memory architectures. Table~\ref{tab:applications} summarizes the spacetime advantages of our work (using simultaneous fan-out) for the applications surveyed in this Section.

\begin{table}[h]
\renewcommand{\arraystretch}{1.3} \centering
\begin{tabular}{P{0.22\textwidth}|P{0.23\textwidth}}
         \textbf{Application} & Spacetime costs \\ \hline
         \textbf{SWAP Test} between two $k = \frac{N-1}{2}$ qubit registers & (0 ancilla for all) \\
         Our work & $14 = O(1)$ depth \\
Serialized & $\sim 14 k = O(N)$ depth \\
Coarse-grained sched. \cite{hoyer2005quantum} &  $\sim 12k = O(N)$ depth \\
Fine-grained sched. \cite{hoyer2005quantum} & $\sim 9k = O(N)$ depth \\ \hline
         \textbf{Hadamard Test}; $N$-qubit circuit; $U$ has depth $D$ & \\
         Our work & $O(D)$ depth, 0 ancilla \\
         Other approaches (Table~\ref{tab:controlled_U_cost}) & $O(ND)$ depth, $O(N^2)$ ancilla, or $\Omega(2^N)$ compile time \\ \hline
         \textbf{Explicit Memory} with $n$ index qubits and bitwidth $W$ \\
         Our work & $O(n)$ depth, 0 ancilla \\
         Bucket-Brigade QRAM \cite{arunachalam2015robustness} & $O(W 2^n)$ depth, 0 ancilla \\
         Parallel QRAM \cite{di2020fault} & $O(W n)$ depth, $O(2^n)$ ancilla \\ \hline
         \textbf{Implicit Memory} with $n$ index qubits and bitwidth $W$ & ($\sim 1 \cdot n$ ancilla for both)\\
         Our work &  $O(2^n)$ depth \\
         QROM \cite{babbush2018encoding} & $O(W 2^n)$ depth
\end{tabular}
\caption{Summary of space (ancilla qubits) and time (depth) costs for different applications. Our work leverages simultaneous fan-out to attain asymptotic advantages.}
\label{tab:applications}
\end{table}

\subsection{SWAP Test} \label{subsec:swap_test}

One of the most important \cite{schuld2018supervised} procedures in quantum computing, especially NISQ machine learning algorithms, is the calculation of inner products between quantum states. This inner product reports the \textit{overlap} or similarity between states. For two qubit registers $\ket{A}$ and $\ket{B}$, this overlap is denoted by $|\braket{A|B}|^2 $. For equal states $|\braket{A|B}|^2 = 1$, and for orthogonal states $|\braket{A|B}|^2 = 0$.

The calculation of this overlap is a procedure known as the SWAP Test. The SWAP Test features heavily in NISQ applications such as quantum kernel classification, which was introduced in \cite{schuld2019quantum} and realized experimentally on IBM's quantum hardware in \cite{havlivcek2019supervised}. These quantum kernel methods are noise resilient and amenable to noise mitigation \cite{havlivcek2019supervised}. Further work \cite{ghobadi2019power} has introduced kernels that have strong complexity theory foundations for hardness of classical simulability. All of these kernel methods require the evaluation of inner product overlaps. The SWAP Test is also integral to cost function evaluation in NISQ-friendly deep quantum neural networks \cite{beer2019efficient}. In the near-term (and in fact current-term), experimental sequences in quantum sensing \cite{zaiser2016enhancing} are essentially overlap measurements.

The SWAP Test has a very simple form. It is essentially just the case of Controlled-$U$ with $U = \text{SWAP}$. First, we examine the decomposition of a SWAP between two qubits:
$$
\Qcircuit @C=0.4em @R=0.5em {
& \multigate{1}{\text{SWAP}} & \qw &&&&&\raisebox{-1.9em}{:=}&&&&&& \Qswap \qwx[1] & \qw &&&&&\raisebox{-1.9em}{=}&&&&&& \Targ     & \Ctrl{1} & \Targ  & \qw   \\
& \ghost{\text{SWAP}}        & \qw &&&&&&&&&&& \Qswap & \qw &&&&&&&&&&& \Ctrl{-1} & \Targ    & \Ctrl{-1} & \qw \\
}
$$
This decomposition is equivalent to the triple XOR sequence for in-place SWAPs of classical bits. For a SWAP Test, we need to perform this $U = \text{SWAP}$ sequence not just between two individual qubits, but between two registers of qubits. Moreover, the SWAP is controlled on an ancilla qubit. The SWAP Test also requires a Hadamard-sandwich around the controls, and a measurement of the ancilla. After executing such a circuit, the overlap between the two registers is related by a simple function to the probability of measuring $\ket{0}$ on the ancilla. Repeated executions can therefore estimate the overlap to a desired precision.

Let us concretely consider the example of a SWAP Test on two two-qubit registers, $\ket{A = A_1 A_0}$ and $\ket{B = B_1 B_0}$. To disambiguate the gates, we have used colors and interleaved the bit ordering of the $\ket{A}$ and $\ket{B}$ registers below:
$$
\Qcircuit @C=0.4em @R=0.6em {
& \lstick{\ket{0}}   & \qw  & \gate{H} & \Ctrl{4}                & \Ctrl{4}               & \Ctrl{4}                & \gate{H} & \meter \\
& \lstick{\ket{A_0}} & \qw  & \qw      & \targ{\color{red}}      & \ctrl{1}{\color{red}}  & \targ{\color{red}}      & \qw      & \qw \\
& \lstick{\ket{B_0}} & \qw  & \qw      & \ctrl{-1}{\color{red}}  & \targ{\color{red}}     & \ctrl{-1}{\color{red}}  & \qw      & \qw \\
& \lstick{\ket{A_1}} & \qw  & \qw      & \targ{\color{blue}}     & \ctrl{1}{\color{blue}} & \targ{\color{blue}}     & \qw      & \qw \\
& \lstick{\ket{B_1}} & \qw  & \qw      & \ctrl{-1}{\color{blue}} & \targ{\color{blue}}    & \ctrl{-1}{\color{blue}} & \qw      & \qw \\
}
$$

Under standard serialization of the shared-control gates, the depth is 63 at best from fine-grained scheduling. However, our Controlled-$U$ synthesis procedure, specifically the shared-control Toffoli's decomposition, is directly applicable here. The resulting SWAP Test depth is $3 \times 12 = 36$ (ignoring the two Hadamard gates). Moreover, our procedure always yields a constant depth of 36 layers regardless of the circuit width $N$, whereas serialized approaches scale as $O(N)$.

While this asymptotic advantage is already appealing, we can attain even further cost reductions to our constants via a circuit identity. It can be shown that the outer two controls on the ancilla qubit can be removed \cite{garcia2013swap, nielsen2002quantum}. After this optimization, the final circuit has a depth of just 14 layers, regardless of the size of the SWAP Test. To illustrate for larger $N$, this \href{https://algassert.com/quirk\#circuit=\%7B\%22cols\%22\%3A\%5B\%5B\%22H\%22\%5D\%2C\%5B1\%2C\%22X\%22\%2C1\%2C1\%2C1\%2C\%22\%E2\%80\%A2\%22\%5D\%2C\%5B1\%2C1\%2C\%22X\%22\%2C1\%2C1\%2C1\%2C\%22\%E2\%80\%A2\%22\%5D\%2C\%5B1\%2C1\%2C1\%2C\%22X\%22\%2C1\%2C1\%2C1\%2C\%22\%E2\%80\%A2\%22\%5D\%2C\%5B1\%2C1\%2C1\%2C1\%2C\%22X\%22\%2C1\%2C1\%2C1\%2C\%22\%E2\%80\%A2\%22\%5D\%2C\%5B1\%2C1\%2C1\%2C1\%2C1\%2C\%22H\%22\%2C\%22H\%22\%2C\%22H\%22\%2C\%22H\%22\%5D\%2C\%5B1\%2C\%22\%E2\%80\%A2\%22\%2C1\%2C1\%2C1\%2C\%22X\%22\%5D\%2C\%5B1\%2C1\%2C\%22\%E2\%80\%A2\%22\%2C1\%2C1\%2C1\%2C\%22X\%22\%5D\%2C\%5B1\%2C1\%2C1\%2C\%22\%E2\%80\%A2\%22\%2C1\%2C1\%2C1\%2C\%22X\%22\%5D\%2C\%5B1\%2C1\%2C1\%2C1\%2C\%22\%E2\%80\%A2\%22\%2C1\%2C1\%2C1\%2C\%22X\%22\%5D\%2C\%5B1\%2C1\%2C1\%2C1\%2C1\%2C\%22Z\%5E-\%C2\%BC\%22\%2C\%22Z\%5E-\%C2\%BC\%22\%2C\%22Z\%5E-\%C2\%BC\%22\%2C\%22Z\%5E-\%C2\%BC\%22\%5D\%2C\%5B\%22\%E2\%80\%A2\%22\%2C1\%2C1\%2C1\%2C1\%2C\%22X\%22\%2C\%22X\%22\%2C\%22X\%22\%2C\%22X\%22\%5D\%2C\%5B1\%2C1\%2C1\%2C1\%2C1\%2C\%22Z\%5E\%C2\%BC\%22\%2C\%22Z\%5E\%C2\%BC\%22\%2C\%22Z\%5E\%C2\%BC\%22\%2C\%22Z\%5E\%C2\%BC\%22\%5D\%2C\%5B1\%2C\%22\%E2\%80\%A2\%22\%2C1\%2C1\%2C1\%2C\%22X\%22\%5D\%2C\%5B1\%2C1\%2C\%22\%E2\%80\%A2\%22\%2C1\%2C1\%2C1\%2C\%22X\%22\%5D\%2C\%5B1\%2C1\%2C1\%2C\%22\%E2\%80\%A2\%22\%2C1\%2C1\%2C1\%2C\%22X\%22\%5D\%2C\%5B1\%2C1\%2C1\%2C1\%2C\%22\%E2\%80\%A2\%22\%2C1\%2C1\%2C1\%2C\%22X\%22\%5D\%2C\%5B1\%2C1\%2C1\%2C1\%2C1\%2C\%22Z\%5E-\%C2\%BC\%22\%2C\%22Z\%5E-\%C2\%BC\%22\%2C\%22Z\%5E-\%C2\%BC\%22\%2C\%22Z\%5E-\%C2\%BC\%22\%5D\%2C\%5B\%22\%E2\%80\%A2\%22\%2C1\%2C1\%2C1\%2C1\%2C\%22X\%22\%2C\%22X\%22\%2C\%22X\%22\%2C\%22X\%22\%5D\%2C\%5B1\%2C\%22Z\%5E\%C2\%BC\%22\%2C\%22Z\%5E\%C2\%BC\%22\%2C\%22Z\%5E\%C2\%BC\%22\%2C\%22Z\%5E\%C2\%BC\%22\%2C\%22Z\%5E\%C2\%BC\%22\%2C\%22Z\%5E\%C2\%BC\%22\%2C\%22Z\%5E\%C2\%BC\%22\%2C\%22Z\%5E\%C2\%BC\%22\%5D\%2C\%5B\%22\%E2\%80\%A2\%22\%2C\%22X\%22\%2C\%22X\%22\%2C\%22X\%22\%2C\%22X\%22\%5D\%2C\%5B1\%2C1\%2C1\%2C1\%2C1\%2C\%22H\%22\%2C\%22H\%22\%2C\%22H\%22\%2C\%22H\%22\%5D\%2C\%5B1\%2C\%22Z\%5E-\%C2\%BC\%22\%2C\%22Z\%5E-\%C2\%BC\%22\%2C\%22Z\%5E-\%C2\%BC\%22\%2C\%22Z\%5E-\%C2\%BC\%22\%5D\%2C\%5B\%22\%E2\%80\%A2\%22\%2C\%22X\%22\%2C\%22X\%22\%2C\%22X\%22\%2C\%22X\%22\%5D\%2C\%5B1\%2C\%22X\%22\%2C1\%2C1\%2C1\%2C\%22\%E2\%80\%A2\%22\%5D\%2C\%5B1\%2C1\%2C\%22X\%22\%2C1\%2C1\%2C1\%2C\%22\%E2\%80\%A2\%22\%5D\%2C\%5B1\%2C1\%2C1\%2C\%22X\%22\%2C1\%2C1\%2C1\%2C\%22\%E2\%80\%A2\%22\%5D\%2C\%5B1\%2C1\%2C1\%2C1\%2C\%22X\%22\%2C1\%2C1\%2C1\%2C\%22\%E2\%80\%A2\%22\%5D\%2C\%5B\%22Z\%22\%5D\%2C\%5B\%22H\%22\%5D\%5D\%7D}{Quirk Link} shows an interactive SWAP Test circuit for computing the overlap of two four-qubit registers, with an ancilla qubit on the top.

% While this asymptotic advantage is already appealing, we can attain even further cost reductions to our constants via a circuit identity. It can be shown that the outer two controls on the ancilla qubit can be removed \cite{garcia2013swap, nielsen2002quantum}. The resulting circuit is below in Figure~\ref{fig:swap_test}.

% \begin{figure}[h]
% $$
% \Qcircuit @C=0.4em @R=0.6em {
% & \lstick{\ket{0}}    & \qw & \gate{H}                & \Ctrl{4}               & \gate{H}                & \meter \\
% & \lstick{\ket{A_0}}  & \qw              & \targ{\color{red}}      & \ctrl{1}{\color{red}}  & \targ{\color{red}}      & \qw \\
% & \lstick{\ket{B_0}}  & \qw              & \ctrl{-1}{\color{red}}  & \targ{\color{red}}     & \ctrl{-1}{\color{red}}  & \qw \\
% & \lstick{\ket{A_1}}  & \qw              & \targ{\color{blue}}     & \ctrl{1}{\color{blue}} & \targ{\color{blue}}     & \qw \\
% & \lstick{\ket{B_1}}  & \qw              & \ctrl{-1}{\color{blue}} & \targ{\color{blue}}    & \ctrl{-1}{\color{blue}} & \qw \\
% }
% $$    \caption{Optimized SWAP Test between two two-qubit registers. The shared-control Toffoli's are executed simultaneously via Figure~\ref{fig:toffoli_parallelization}, leading to a depth of 14 layers. The depth is independent of $N$.}
%     \label{fig:swap_test}
% \end{figure}

% This final circuit has a depth of just 14 layers, regardless of the size of the SWAP Test. To illustrate for larger $N$, this \href{https://bit.ly/2QFw3Xf}{Quirk Link} shows an interactive SWAP Test circuit for computing the overlap of two-four qubit registers, with an ancilla qubit on the top.

\subsubsection{Interference Circuit}
\begin{figure}[h]
$$
\Qcircuit @C=1em @R=0.5em {
& \qw & \gate{H} & \Ctrl{1}           & \ctrlo{1}          & \gate{H} & \meter \\
& \qw & \qw      & \multigate{3}{U_A} & \multigate{3}{U_B} & \qw & \qw     \\
& \qw & \qw      & \ghost{U_A}        & \ghost{U_B}        & \qw & \qw    \\
& & \raisebox{0.5em}{\vdots} & & & \raisebox{0.5em}{\vdots} \\
& \qw & \qw      & \ghost{U_A}        & \ghost{U_B}        & \qw & \qw \\
}
$$
    \caption{The interference circuit computes the overlap between $k$-qubit states, $\ket{A}$ and $\ket{B}$, with just $k+1$ qubits.}
    \label{fig:interference_circuit}
\end{figure}
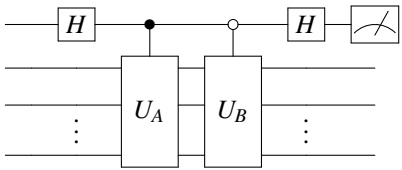

Recent work has explored alternatives to the traditional SWAP Test, with the aim of reducing spacetime costs. The most promising one is the interference circuit \cite{schuld2017implementing, schuld2018supervised}, which halves the qubit width requirement. Whereas the traditional SWAP Test requires $2k+1$ qubits to compute the overlap of two $k$-qubit registers, the interference circuit only requires $k + 1$ qubits. In order to use the interference circuit, we must know the sequences of gates $U_A$ and $U_B$ that can create $\ket{A}$ and $\ket{B}$, respectively. In practice, this is indeed the case for useful applications. The interference circuit has the following simple form shown in Figure~\ref{fig:interference_circuit}. As in the traditional SWAP Test, the overlap is a simple function of the probability of measuring $\ket{0}$ on the ancilla.

% \subsection{Alternatives to Traditional SWAP Test}
% Recent work has explored alternatives to the traditional SWAP Test, with the aim of reducing spacetime costs. The first is the interference circuit \cite{schuld2017implementing, schuld2018supervised}, which halves the qubit width requirement. Whereas the traditional SWAP Test requires $2k+1$ qubits to compute the overlap of two $k$-qubit registers, the interference circuit only requires $k + 1$ qubits. In order to use the interference circuit, we must know the sequences of gates $U_A$ and $U_B$ that can create $\ket{A}$ and $\ket{B}$, respectively. In practice, this is indeed the case for useful applications. The interference circuit has the following simple form shown in Figure~\ref{fig:interference_circuit}. As in the traditional SWAP Test, the overlap is a simple function of the probability of measuring $\ket{0}$ on the ancilla.

The open-control (open circle) on $U_B$ activates on $\ket{0}$ and can therefore be replaced with an ordinary control surrounded by NOT ($\oplus$) gates. Therefore our Controlled-$U$ is directly applicable to the interference circuit, and it allows overlap calculation with no asymptotic depth overhead relative to $U_A$ and $U_B$. This is again a linear $O(N)$ speedup via fan-out.

\subsection{Hadamard Test} \label{subsec:hadamard_test}

\begin{figure}
    \centering
$$
\Qcircuit @C=1em @R=0.5em {
& \gate{H} & \Ctrl{1} & \gate{H} & \meter \\
& \qw      & \multigate{3}{U} & \qw & \qw \\
& \qw      & \ghost{U} & \qw  & \qw \\
& \raisebox{0.5em}{\vdots}   &           & \raisebox{0.5em}{\vdots} \\
& \qw      & \ghost{U} & \qw  & \qw \\
}
$$    \caption{Circuit for the Hadamard Test.}
    \label{fig:hadamard_test}
\end{figure}
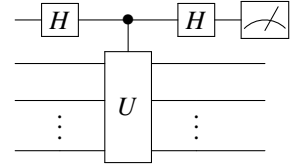

The SWAP Test is a specific case of a more general procedure called the Hadamard Test. The Hadamard Test has a very simple and familiar form shown in Figure~\ref{fig:hadamard_test}. This is essentially just the Controlled-$U$ operation we focused on in Section~\ref{sec:universality}. Moreover, the SWAP Test is just the case where $U = \text{SWAP}$. Selecting other $U$ makes the Hadamard Test give rise to a wide variety of applications. We list our benchmarked applications in Table~\ref{tab:hadamard_test}. %The specific applications and the corresponding implementations of $U$ are documented in \texttt{benchmarks.ipynb}. 
There are numerous additional applications of the Hadamard Test, such as training Quantum Boltzmann Machines \cite{wiebe2019generative}, gradient evaluation \cite{schuld2019evaluating, guerreschi2017practical, mitarai2019methodology, microsoft2020estimategradient}, and Jones polynomial approximation \cite{aharonov2009polynomial}. 

\begin{table}[h]
\small
\renewcommand{\arraystretch}{1.4} \centering
\begin{tabular}{P{0.18\textwidth}|p{0.25\textwidth}}
Application         & Description          \\ \hline
Variational Quantum Linear Solver \cite{bravo2019variational, xu2019variational, huang2019near} & Algorithm for solver large linear systems using NISQ hardware \\
Matrix elements of group representation \cite{jordan2008fast, coles2018quantum} & Group theory problem; $U$ is essentially the Quantum Fourier Transform \\
Entanglement spectroscopy \cite{johri2017entanglement} & Computation of entanglement spectrum of arbitrary quantum states \\
Controlled Density Matrix Exponentiation (DME) \cite{kjaergaard2020quantum} & Several appliations, e.g. for private quantum software \cite{marvian2016universal} \\
% Quantum Boltzmann Machines \cite{wiebe2019generative} & Representationally-rich neural network; Hadamard test used in training \\
% Gradient evaluation \cite{schuld2019evaluating, guerreschi2017practical, mitarai2019methodology, microsoft2020estimategradient} & Computation of partial derivatives w.r.t circuit parameters; needed for variational algorithms \\
% Approximating Jones polynomials \cite{aharonov2009polynomial} & Important for topological quantum field theory \\
\end{tabular}
\caption{Applications of the Hadamard Test. Each corresponds to a different choice of $U$.}
\label{tab:hadamard_test}
\end{table}

% \section{Steane Code Measurement}
% See Fig 10.16 in Mike and Ike. Steane code is \cite{steane1996multiple}.

% \section{DQC1 Kernel}
% DQC1 Power of One Qubit for ML Kernel has one control.

\subsection{Quantum Memory Architectures}
Next, we investigate the use of fan-out to improve the implementation of quantum memory, which speeds up or enables many quantum algorithms \cite{wiebe2014quantum, verdon2017quantum}. The high-level function of a quantum memory is similar to that of a classical memory: $n$ index bits enumerate over $2^n$ memory cells. Following the notation of \cite{arad2010quantum}, we denote the $n$ index bits as the $\ket{b}$ register and the $2^n$ memory cells as the $\ket{m}$ register. As in the classical case, we expect that setting the index register to $\ket{i}$ should allow us to retrieve the $i$th memory cell, $\ket{m_i}$. However, for a quantum memory, we also require the retrieval to work over superpositions of index qubits. For example, setting $\ket{b}$ to $\frac{1}{\sqrt{2}}[\ket{000} + \ket{111}]$ should retrieve the superposition, $\frac{1}{\sqrt{2}}[\ket{m_0} + \ket{m_7}]$.

In this section, we apply the fan-out primitive to both explicit and implicit quantum memories, which we define below. We demonstrate considerable improvements---exponential and linear respectively---over prior work, as summarized in Table~\ref{tab:applications}. These improvements are important because the cost of quantum memory is often the principal bottleneck for realizing practical speedups. While it remains unclear if quantum memory architectures will be feasible \cite{aaronson2015read, arunachalam2015robustness, biamonte2017quantum, preskill2018quantum} even for future fault-tolerant devices, our proposed improvements at least justify a re-assessment of the feasibility.

\subsubsection{Explicit Quantum Memory}
In an explicit quantum memory, the $2^n$ memory cells are each explicitly stored in qubit registers. In this sense, an explicit quantum memory is akin to a $2^n$--to--1 multiplexer or data selector from classical electronics. As discussed, the quantum variant should extend to the case where select lines are in superposition. Moreover, each of the $2^n$ memory cells is stored in a qubit register, so each memory cell can itself contain a quantum (superposition) state.

The dominant architecture for this explicit quantum memory is termed Quantum Random Access Memory. The bucket brigade design of QRAM was introduced in \cite{giovannetti2008quantum, giovannetti2008architectures} and cast to the quantum circuit model in \cite{arunachalam2015robustness}. This bucket brigade QRAM requires $\sim 2 \cdot 2^n$ qubits and $O(W2^n)$ depth. Later work \cite{di2020fault} was able to parallelize execution to achieve $O(Wn)$ depth, but requires an additional $\sim 6 \cdot 2^n$ ancilla qubits. We now present a novel architecture for explicit quantum memory that requires only $O(n)$ depth, with 0 ancilla qubits.

\begin{figure}
$$
\Qcircuit @C=1.5em @R=1em {
& \lstick{\ket{{b}_0}} & \qw                          & \qw                  & \Ctrl{4} & \qw            &  \Ctrl{4} & \qw                  & \qw                          & \qw \\
& \lstick{\ket{{b}_1}} & \qw                          & \Ctrl{5}             & \qw      & \qw            &  \qw      & \Ctrl{5}             & \qw                          & \qw \\
& \lstick{\ket{{b}_2}} & \Ctrl{8}                     & \qw                  & \qw      & \qw            &  \qw      & \qw                  & \Ctrl{8}                     & \qw \\
& \lstick{\ket{{m}_0}} & \qswap{\color{red!100!blue}} & \qswap{\color{red}}  & \Qswap   & \Qswap \qwx[8] &  \Qswap   & \qswap{\color{red}}  & \qswap{\color{red!100!blue}} & \qw \\
& \lstick{\ket{{m}_1}} & \qswap{\color{red!66!blue}}  & \qswap{\color{blue}} & \Qswap   & \qw            &  \Qswap   & \qswap{\color{blue}} & \qswap{\color{red!66!blue}}  & \qw \\
& \lstick{\ket{{m}_2}} & \qswap{\color{red!33!blue}}  & \qswap{\color{red}}  & \qw      & \qw            &  \qw      & \qswap{\color{red}}  & \qswap{\color{red!33!blue}}  & \qw \\
& \lstick{\ket{{m}_3}} & \qswap{\color{red!0!blue}}   & \qswap{\color{blue}} & \qw      & \qw            &  \qw      & \qswap{\color{blue}} & \qswap{\color{red!0!blue}}   & \qw \\
& \lstick{\ket{{m}_4}} & \qswap{\color{red!100!blue}} & \qw                  & \qw      & \qw            &  \qw      & \qw                  & \qswap{\color{red!100!blue}} & \qw \\
& \lstick{\ket{{m}_5}} & \qswap{\color{red!66!blue}}  & \qw                  & \qw      & \qw            &  \qw      & \qw                  & \qswap{\color{red!66!blue}}  & \qw \\
& \lstick{\ket{{m}_6}} & \qswap{\color{red!33!blue}}  & \qw                  & \qw      & \qw            &  \qw      & \qw                  & \qswap{\color{red!33!blue}}  & \qw \\
& \lstick{\ket{{m}_7}} & \qswap{\color{red!0!blue}}   & \qw                  & \qw      & \qw            &  \qw      & \qw                  & \qswap{\color{red!0!blue}}   & \qw \\
& \lstick{\ket{\text{load/store}}} & \qw                          & \qw                  & \qw      & \Qswap         &  \qw      & \qw                  & \qw                          & \qw \\
}
$$
    \caption{Architecture for an explicit quantum memory with $n = 3$ index qubits and $2^n = 8$ memory cells of bitwidth $W = 1$. \href{https://algassert.com/quirk\#circuit=\%7B\%22cols\%22\%3A\%5B\%5B\%22Counting3\%22\%2C1\%2C1\%2C\%7B\%22id\%22\%3A\%22Ryft\%22\%2C\%22arg\%22\%3A\%220*pi\%2F7\%22\%7D\%2C\%7B\%22id\%22\%3A\%22Ryft\%22\%2C\%22arg\%22\%3A\%221*pi\%2F7\%22\%7D\%2C\%7B\%22id\%22\%3A\%22Ryft\%22\%2C\%22arg\%22\%3A\%222*pi\%2F7\%22\%7D\%2C\%7B\%22id\%22\%3A\%22Ryft\%22\%2C\%22arg\%22\%3A\%223*pi\%2F7\%22\%7D\%2C\%7B\%22id\%22\%3A\%22Ryft\%22\%2C\%22arg\%22\%3A\%224*pi\%2F7\%22\%7D\%2C\%7B\%22id\%22\%3A\%22Ryft\%22\%2C\%22arg\%22\%3A\%225*pi\%2F7\%22\%7D\%2C\%7B\%22id\%22\%3A\%22Ryft\%22\%2C\%22arg\%22\%3A\%226*pi\%2F7\%22\%7D\%2C\%7B\%22id\%22\%3A\%22Ryft\%22\%2C\%22arg\%22\%3A\%227*pi\%2F7\%22\%7D\%5D\%2C\%5B\%5D\%2C\%5B1\%2C1\%2C1\%2C\%22Bloch\%22\%2C\%22Bloch\%22\%2C\%22Bloch\%22\%2C\%22Bloch\%22\%2C\%22Bloch\%22\%2C\%22Bloch\%22\%2C\%22Bloch\%22\%2C\%22Bloch\%22\%5D\%2C\%5B1\%2C1\%2C\%22\%E2\%80\%A2\%22\%2C1\%2C1\%2C1\%2C\%22Swap\%22\%2C1\%2C1\%2C1\%2C\%22Swap\%22\%5D\%2C\%5B1\%2C1\%2C\%22\%E2\%80\%A2\%22\%2C1\%2C1\%2C\%22Swap\%22\%2C1\%2C1\%2C1\%2C\%22Swap\%22\%5D\%2C\%5B1\%2C1\%2C\%22\%E2\%80\%A2\%22\%2C1\%2C\%22Swap\%22\%2C1\%2C1\%2C1\%2C\%22Swap\%22\%5D\%2C\%5B1\%2C1\%2C\%22\%E2\%80\%A2\%22\%2C\%22Swap\%22\%2C1\%2C1\%2C1\%2C\%22Swap\%22\%5D\%2C\%5B1\%2C\%22\%E2\%80\%A2\%22\%2C1\%2C1\%2C\%22Swap\%22\%2C1\%2C\%22Swap\%22\%5D\%2C\%5B1\%2C\%22\%E2\%80\%A2\%22\%2C1\%2C\%22Swap\%22\%2C1\%2C\%22Swap\%22\%5D\%2C\%5B\%22\%E2\%80\%A2\%22\%2C1\%2C1\%2C\%22Swap\%22\%2C\%22Swap\%22\%5D\%2C\%5B1\%2C1\%2C1\%2C\%22Swap\%22\%2C1\%2C1\%2C1\%2C1\%2C1\%2C1\%2C1\%2C\%22Swap\%22\%5D\%2C\%5B\%22\%E2\%80\%A2\%22\%2C1\%2C1\%2C\%22Swap\%22\%2C\%22Swap\%22\%5D\%2C\%5B1\%2C\%22\%E2\%80\%A2\%22\%2C1\%2C\%22Swap\%22\%2C1\%2C\%22Swap\%22\%5D\%2C\%5B1\%2C\%22\%E2\%80\%A2\%22\%2C1\%2C1\%2C\%22Swap\%22\%2C1\%2C\%22Swap\%22\%5D\%2C\%5B1\%2C1\%2C\%22\%E2\%80\%A2\%22\%2C\%22Swap\%22\%2C1\%2C1\%2C1\%2C\%22Swap\%22\%5D\%2C\%5B1\%2C1\%2C\%22\%E2\%80\%A2\%22\%2C1\%2C\%22Swap\%22\%2C1\%2C1\%2C1\%2C\%22Swap\%22\%5D\%2C\%5B1\%2C1\%2C\%22\%E2\%80\%A2\%22\%2C1\%2C1\%2C\%22Swap\%22\%2C1\%2C1\%2C1\%2C\%22Swap\%22\%5D\%2C\%5B1\%2C1\%2C\%22\%E2\%80\%A2\%22\%2C1\%2C1\%2C1\%2C\%22Swap\%22\%2C1\%2C1\%2C1\%2C\%22Swap\%22\%5D\%5D\%7D}{Quirk demo}.} \label{fig:explicit_memory}
\end{figure}
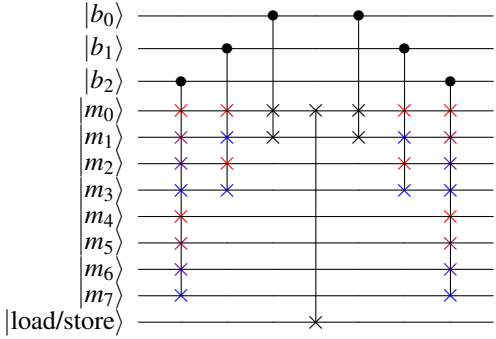

Figure~\ref{fig:explicit_memory} shows our architecture for $n=3$ index qubits. There are $2^3=8$ explicit memory cells, each of single-qubit bitwidth $W=1$. At a high level, the circuit performs a ``migration'' of the target memory cell into $\ket{m_0}$. Consider for example $\ket{\vec{b} = 101}$, which should access $\ket{{m}_5}$. The control on the MSB performs a SWAP between $\ket{{m}_{7654}}$ and $\ket{{m}_{3210}}$, moving $\ket{{m}_5}$ into $\ket{{m}_1}$. The control on the middle index does not activate, but the control on the LSB is activated and SWAPs $\ket{{m}_1}$ into the $\ket{{m}_0}$ destination. Finally, this qubit is swapped into the $\ket{\text{load/store}}$ register. The right half of the circuit reverses the earlier migrations, restoring the other memory cells to their original locations.

The efficiency of this architecture is enabled by the simultaneous execution of controlled SWAPs, which in turn is enabled by the fan-out primitive. As a result, the circuit depth is only $O(n)$. Moreover, while our example shows the $W=1$ bitwidth case, it is apparent that with simultaneous fan-out, $W$ is irrelevant to depth. By contrast, serialization would impose an additional linearity in $W$.

During the preparation of this paper, another proposal was published for $O(n)$-depth and ancilla-free explicit quantum memory \cite{paler2020constant}, which matches our asymptotic costs.

% Deep Restricted (classical) Boltzmann Machine can be trained with quadratically fewer training data accesses if we have QRAM \cite{wiebe2014quantum}. Maybe also in \cite{verdon2017quantum}.

\subsubsection{Implicit Quantum Memory}
Next we consider implicit quantum memory. In this model, the $2^n$ memory cells represent classical (non-superposition) data that is known in advance. In such a case, there is no need to waste qubits to represent the classically-known memory cells. Instead, the memory can be stored implicitly through the classical control, a memory architecture that has been referred to as Quantum Read Only Memory \cite{babbush2018encoding}.

\begin{figure}[h]
$$
\Qcircuit @C=3.1em @R=0.5em {
& \lstick{\ket{{b}_0}} & \ctrlo{1} & \Ctrl{1}  & \ctrlo{1} & \Ctrl{1} & \qw \\
& \lstick{\ket{{b}_1}} & \ctrlo{2} & \ctrlo{2} & \Ctrl{3}  & \Ctrl{3} & \qw \\
&    & \qw     & \Targ       & \Targ     & \Targ    & \qw \\
&  \lstick{\ket{\vec{m}}}  & \Targ     & \Targ       & \qw     & \Targ    & \qw \\
&    & \qw     & \qw       & \Targ     & \Targ    & \qw \gategroup{3}{2}{5}{2}{.5em}{\{} \\
&                               & \vec{m} = 2   & \vec{m} = 3 & \vec{m} = 5 & \vec{m} = 7  \\
}
$$
\caption{Implicit memory storing the first four prime numbers. The $W=3$ bitwidth memory is implicitly defined through classical control, based on the pattern of $\oplus$'s. For anticipated applications, $W$ can be large.}
\label{fig:implicit}
\end{figure}
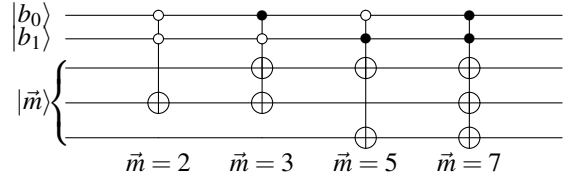

Figure~\ref{fig:implicit} shows an example implicit memory storing the first four prime numbers: $\{00\to 2, 01\to 3, 10\to 5, 11\to 7\}$. The resulting circuit has a simple form, enumerating all $2^n$ indices and associating each index with a corresponding pattern of $\oplus$ gates. Without fan-out, implicit memory has $O(W 2^n)$ depth via the unary iteration optimization in \cite{babbush2018encoding}. However, simultaneous fan-out obviates the scaling in $W$. This is appealing, because for datasets such as images, the bitwidth ($W$) of each record exceeds the number of records.

% \subsection{Potential NISQ Applications of Quantum Memories}

% For explicit quantum memory, good discussion in [Bendetti et al. 2019]

% "PQC models can also handle inputs and outputs that
% are inherently quantum mechanical, i.e., already in superposition. These are often referred to as quantum
% data [67]. Quantum input data could originate remotely,
% for example, from other quantum computers transmitting over a quantum Internet [68]. Otherwise, if a preparation recipe is available, one could prepare the input
% data locally using a suitable encoder circuit. Assuming
% this data preparation is efficient, one can extend supervised and unsupervised learning to quantum states and
% quantum informatio
\section{Technology Modeling: \\ Trapped Ion} \label{sec:trapped_ion}
In this section, we model the implementation of fan-out on trapped ion quantum computers. Trapped ions feature long qubit coherence times~\cite{wang2017single} and gate fidelities exceeding 99.99\% and 99.9\% for single- and two- qubit gates on current hardware \cite{brown2011single, gaebler2016high}. Furthermore, all $N$ qubits can be simultaneously entangled via a global interaction known as the Global Mølmer–Sørensen (GMS) gate \cite{molmer1999multiparticle, sorensen1999quantum}. Recent work \cite{bermudez2017assessing, maslov2018use} has explicitly demonstrated how GMS is essentially equivalent to simultaneous fan-out. Moreover, in the past year, experimental work has merged demonstrating pulse shaping for global interactions \cite{lu2019global, figgatt2019parallel, grzesiak2019efficient} to support the use of GMS both for fan-out and for parallel two-qubit gates on disjoint qubits. Our focus here is on studying differences in speed and fidelity between simultaneous fan-out versus $N-1$ serialized CNOTs. For brevity and to maintain a focus on architectural themes, we omit many physical implementation details here. %These details are documented in \texttt{Trapped\_Ion\_Simulation.ipynb}.

Regarding the potential speedup, \cite{bermudez2017assessing, spiteri2018quantum, grzesiak2019efficient} assert that simultaneous fan-out via GMS is indeed linearly faster than serialized CNOTs. To evaluate the fidelity impact, we performed numerical simulations of fan-out via GMS for $N=2$ to $N=8$ qubits. We constructed a realistic error model that accounts for two sources of noise: overrotation and laser dephasing. Overrotation occurs due to the fact that the angle $\theta$ of the Mølmer-Sørensen rotation is sensitive to motional frequency drifts, and it has higher-order dependence on the motional states~\cite{wang2020high, wu2018noise, debroy2018stabilizer}. An overrotation error can be modeled by replacing $\theta$ by $(1 + \epsilon)\theta$, where $\epsilon$ denotes the overrotation rate.
Laser dephasing results from fluctuations of the optical path length \cite{lee2005phase, wu2018noise, wang2020high}. 

\begin{figure}[h!]
    \centering
    \includegraphics[width=0.45\textwidth]{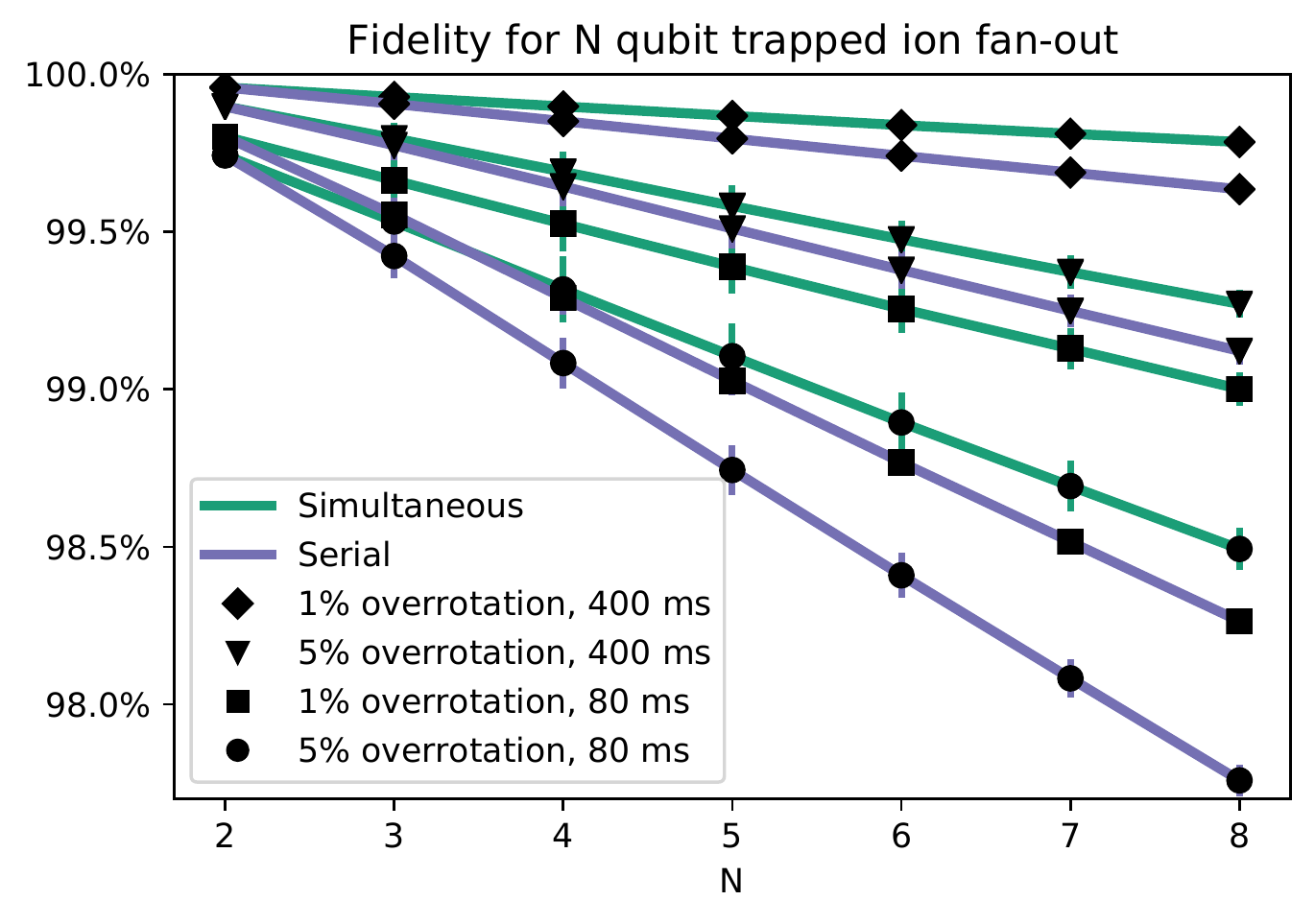}
    \caption{Simulation results for fan-out on trapped ion hardware. Sensitivity analysis performed under four \{overrotation rate, laser coherence time\} scenarios. For each scenario, we simulated fidelity for simultaneous versus serial. Results averaged across 100k stochastic runs per scenario, executed with 50k CPU-core hours on a large computing cluster.}
    \label{fig:trapped_ion_simulation}
\end{figure}

For current trapped ion hardware, we conservatively estimate typical overrotation rates of 5\%. We modeled GMS interaction times of 100 $\mu\text{s}$ \cite{bermudez2017assessing}, contrasted against 80 ms laser coherence time \cite{wang2020high}. To evaluate the sensitivity of our results to these parameters, we also modeled under three future scenarios: 5x lower overrotation rate, 5x longer laser coherence, and both improvements. Our simulations were performed using master-equation simulation in QuTiP \cite{johansson2013qutip}. We performed stochastic simulation over 100k runs per scenario. The fidelity results are shown in Figure~\ref{fig:trapped_ion_simulation}.

Conceptually, overrotation errors affect simultaneous and serial equally. Meanwhile, laser dephasing affects serial more adversely, because the laser dephasing effect on the control qubit accumulates over the additional time required for $N-1$ consecutive CNOTs. Although simultaneous always outperforms serial on our simulations, the exact fidelity advantage is dependent on the parameter settings. For current technology ($\bullet$), simultaneous has an almost 1\% higher fidelity for $N=8$. For the scenario with 5x longer laser coherence ($\blacktriangledown$), simultaneous has almost no fidelity advantage over serial. For the scenario with 5x lower overrotation ($\blacksquare$), simultaneous again has a nearly 1\% fidelity advantage over serial. Also, across all scenarios, the advantage of simultaneous fan-out increases for larger $N$, which is encouraging. While our simulation results are based on a realistic noise model, experimental evaluation is necessary to conclude any definitive fidelity advantage. As cloud access to trapped ion hardware emerges over the coming year, it will be possible to experimentally test these simulated results.

% methodology before results?

\section{Results} \label{sec:results}

\subsection{Methodology}
\begin{figure*}[t!]
\centering
\begin{subfigure}[b]{0.24\textwidth}
\includegraphics[width=\textwidth]{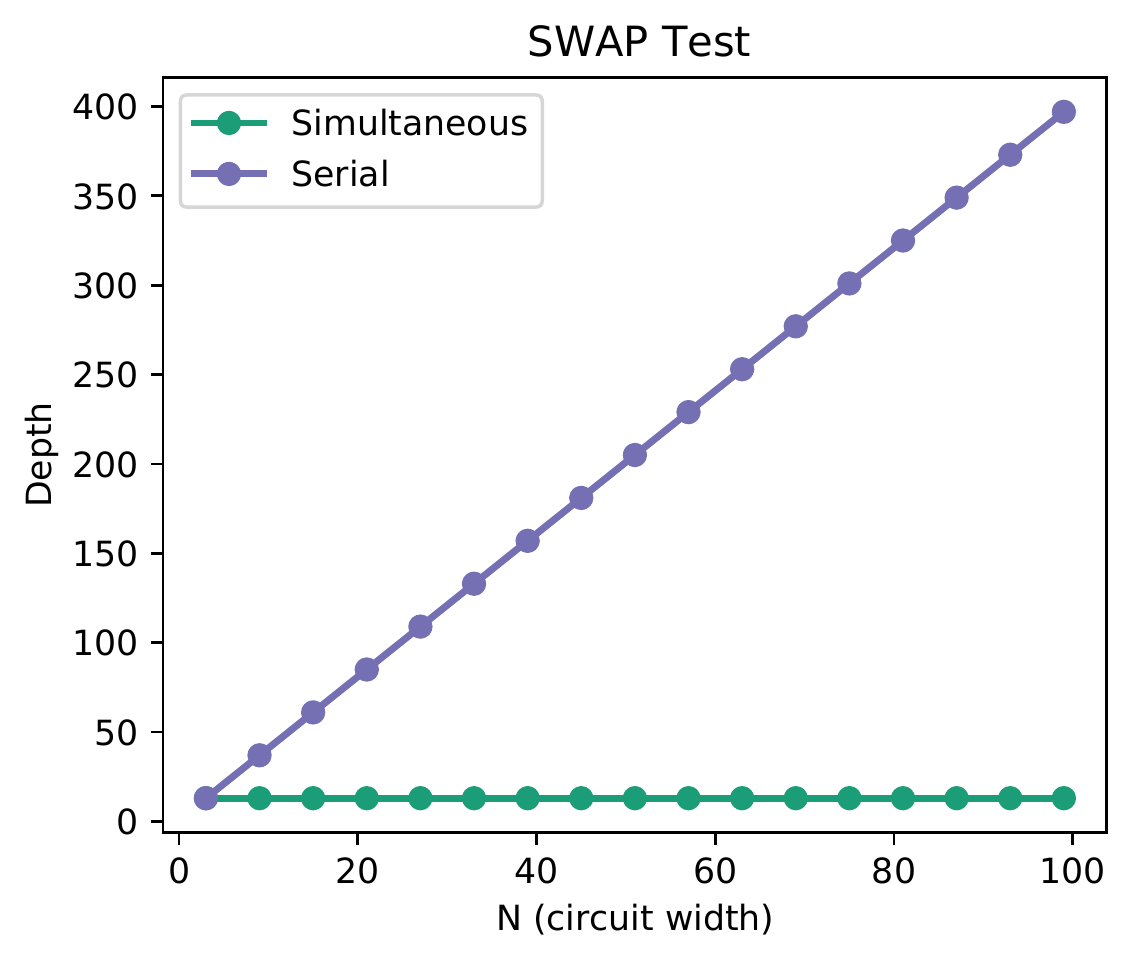}
\end{subfigure}
\hfill
\begin{subfigure}[b]{0.24\textwidth}
\includegraphics[width=\textwidth]{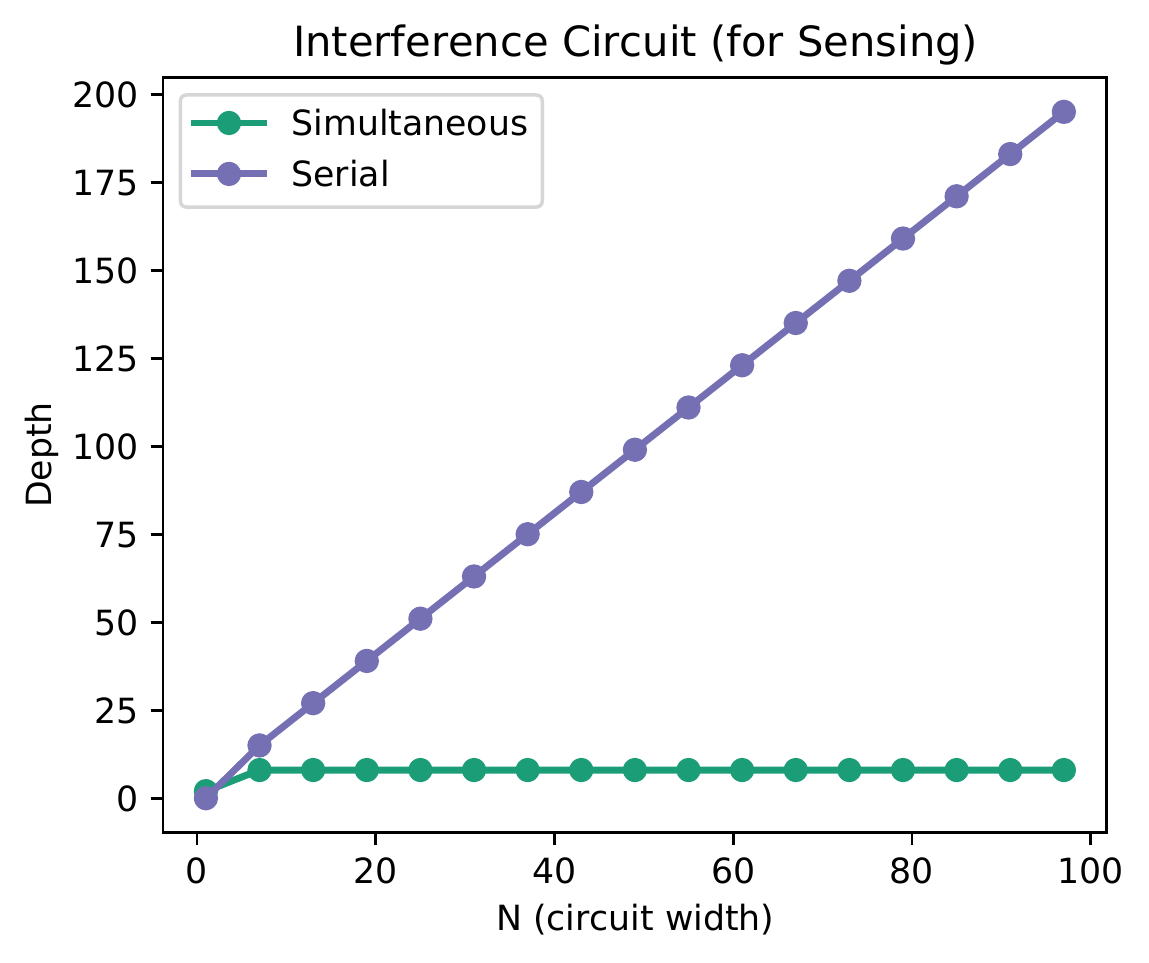}
\end{subfigure}
\hfill
\begin{subfigure}[b]{0.24\textwidth}
\includegraphics[width=\textwidth]{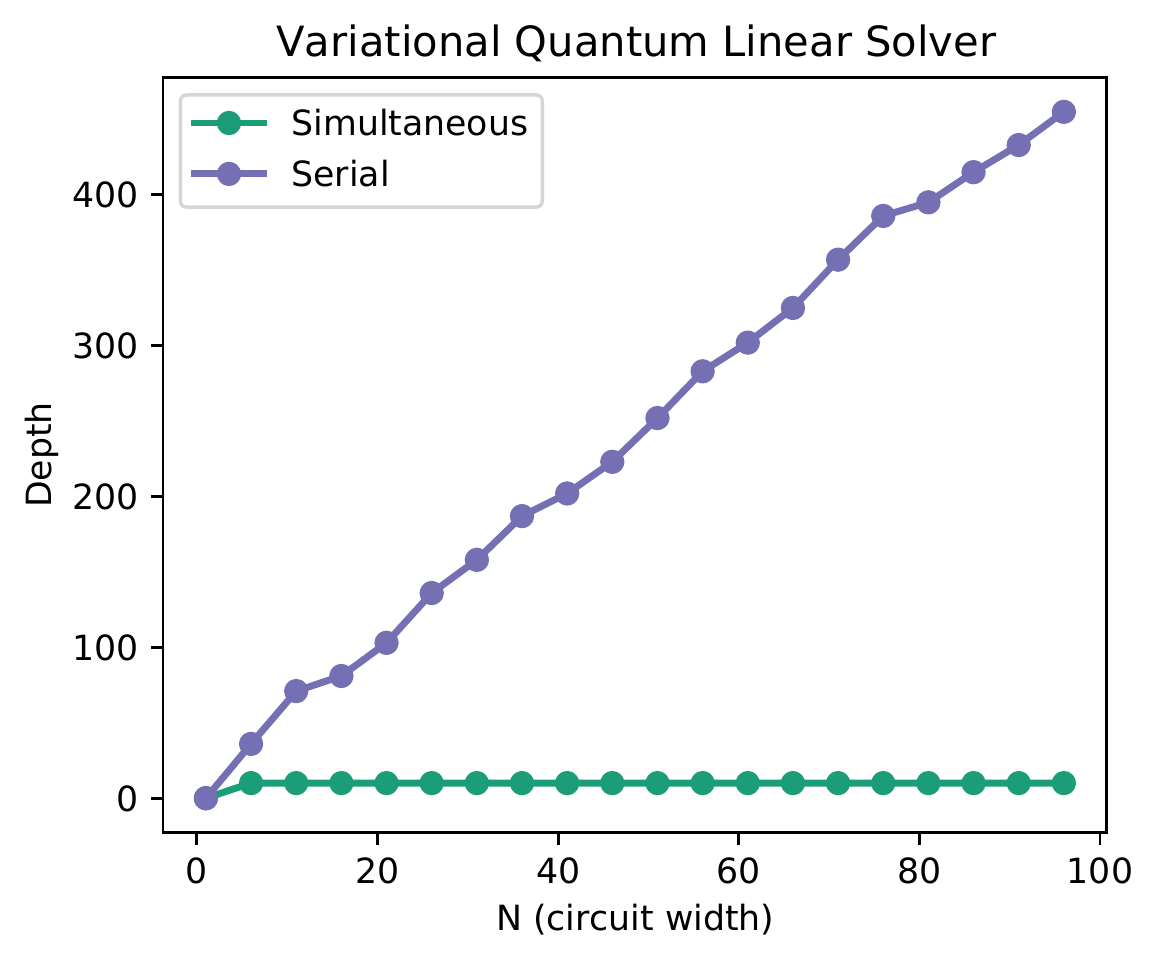}
\end{subfigure}
\hfill
\begin{subfigure}[b]{0.24\textwidth}
\includegraphics[width=\textwidth]{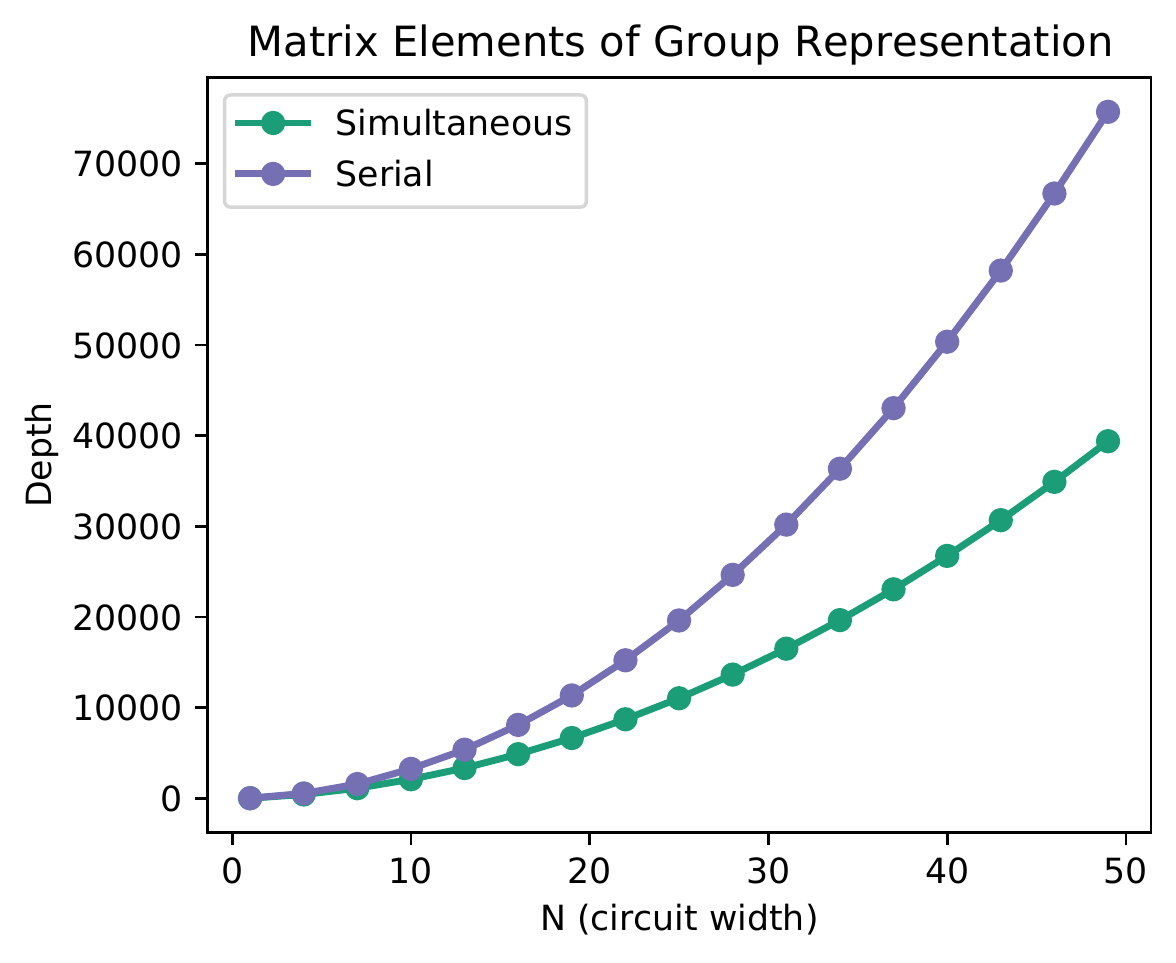}
\end{subfigure}
\hfill
\begin{subfigure}[b]{0.24\textwidth}
\includegraphics[width=\textwidth]{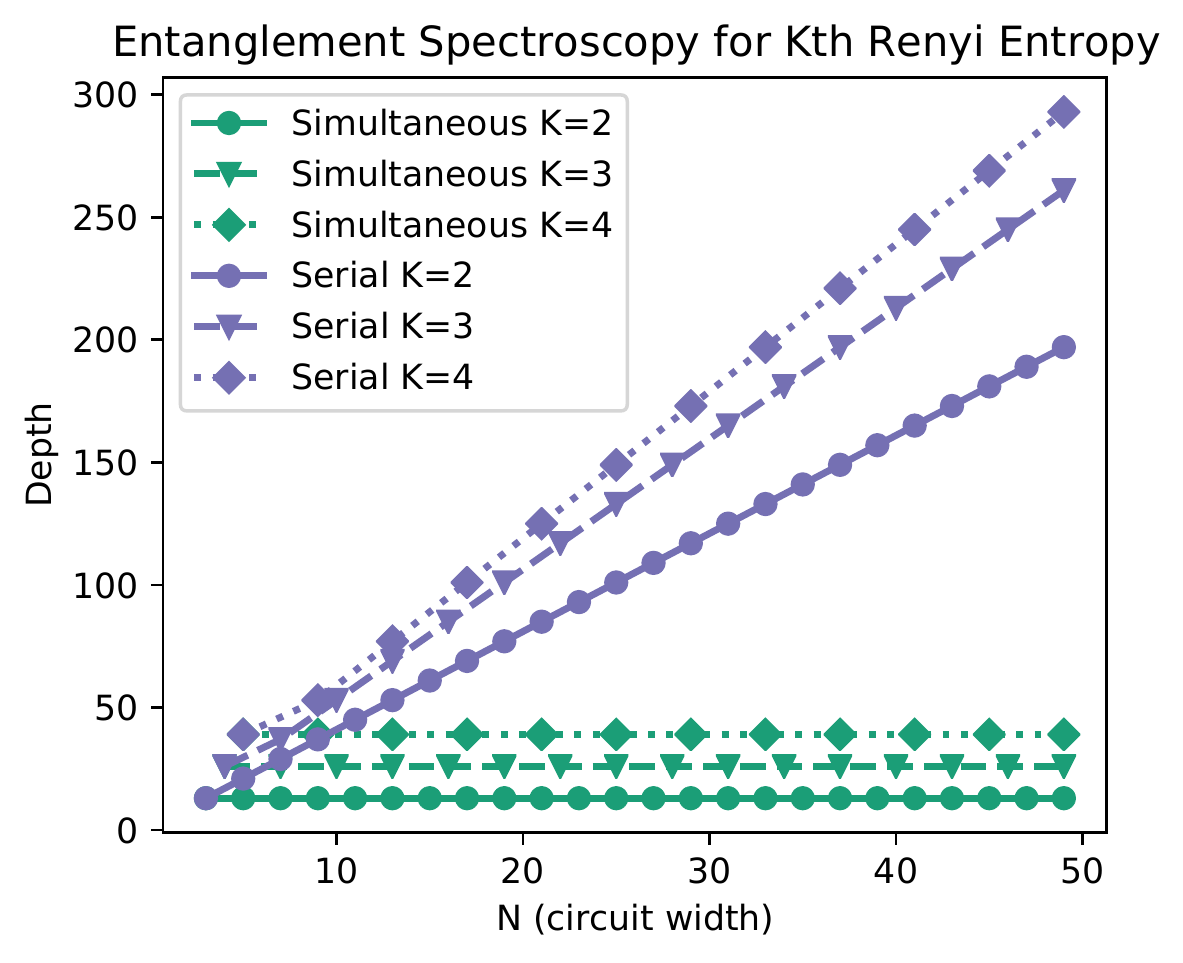}
\end{subfigure}
\hfill
\begin{subfigure}[b]{0.24\textwidth}
\includegraphics[width=\textwidth]{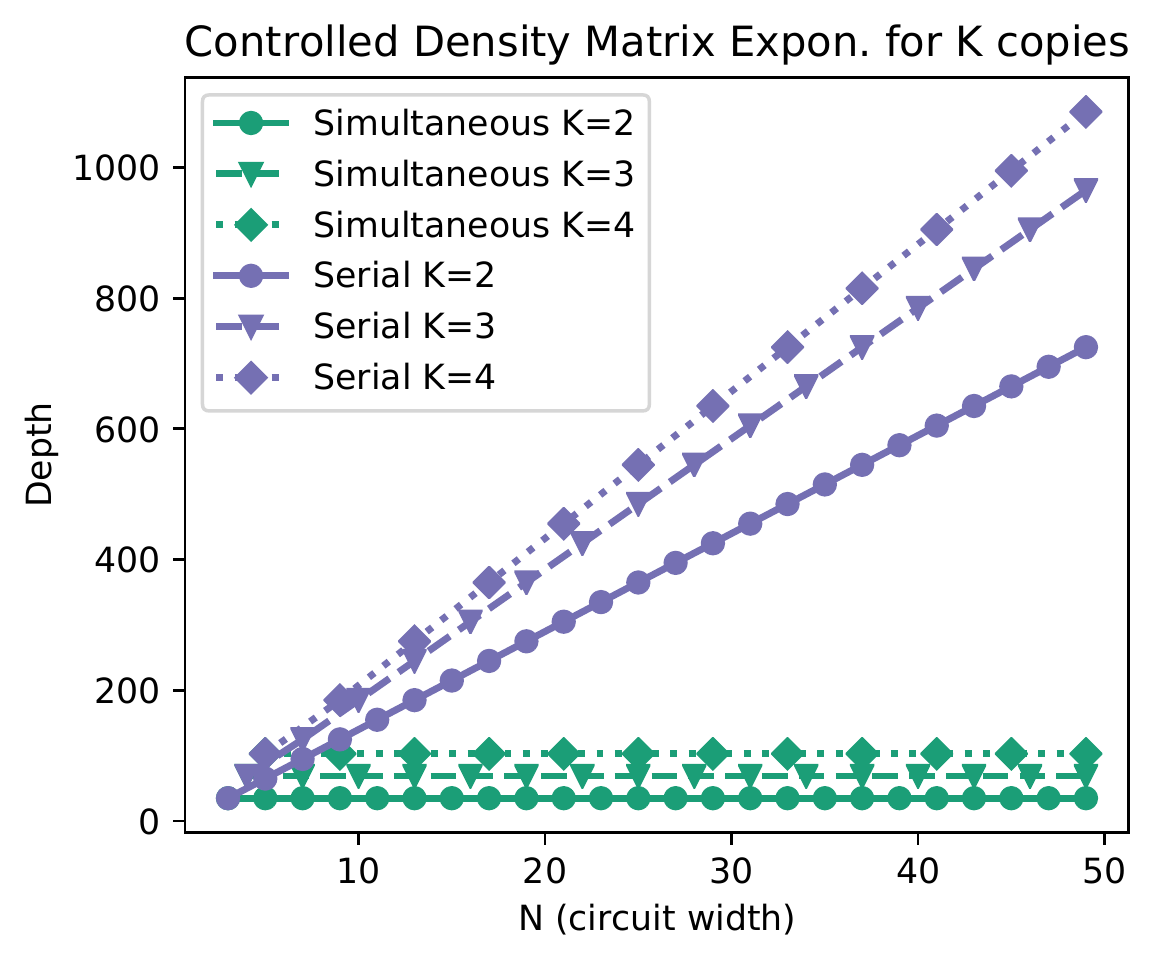}
\end{subfigure}
\hfill
\begin{subfigure}[b]{0.24\textwidth}
\includegraphics[width=\textwidth]{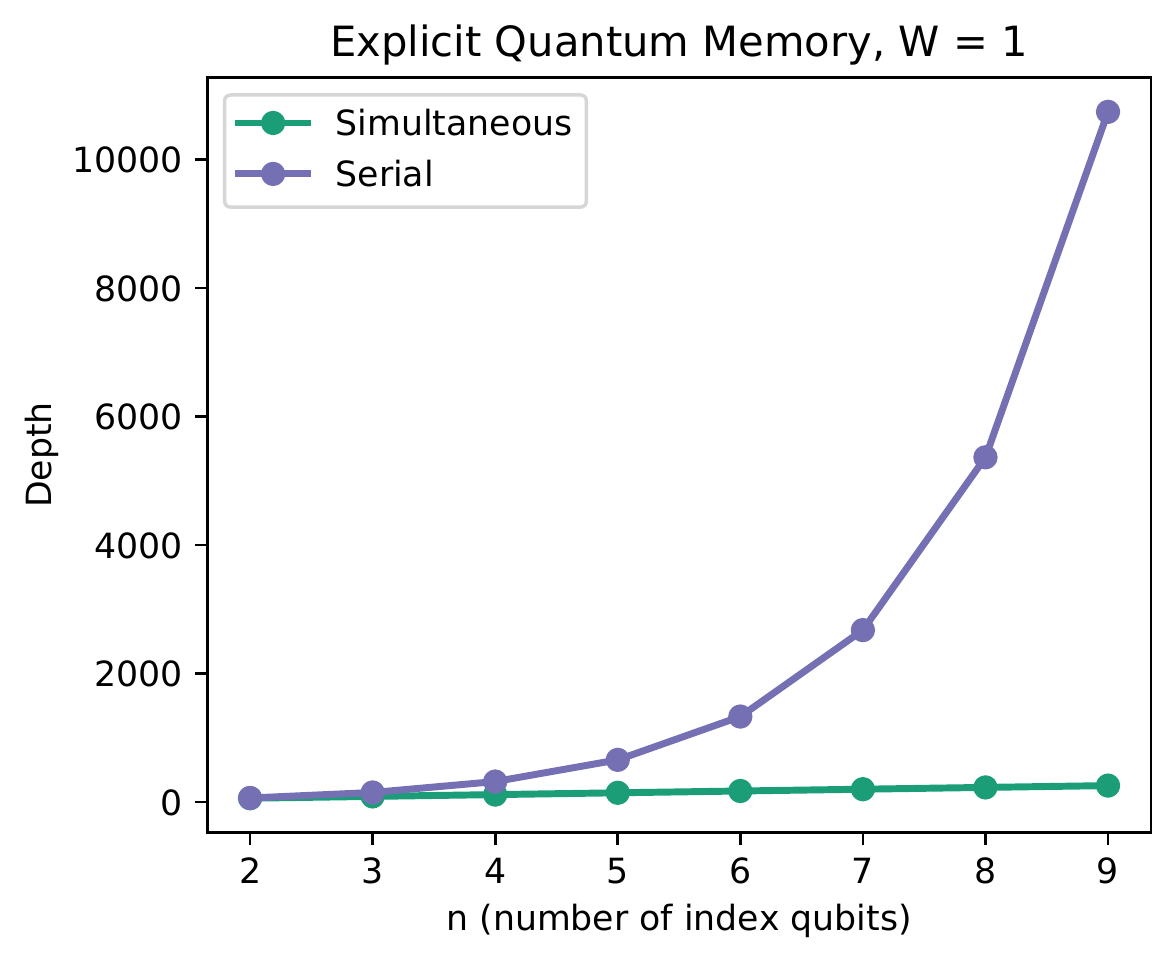}
\end{subfigure}
\hfill
\begin{subfigure}[b]{0.24\textwidth}
\includegraphics[width=\textwidth]{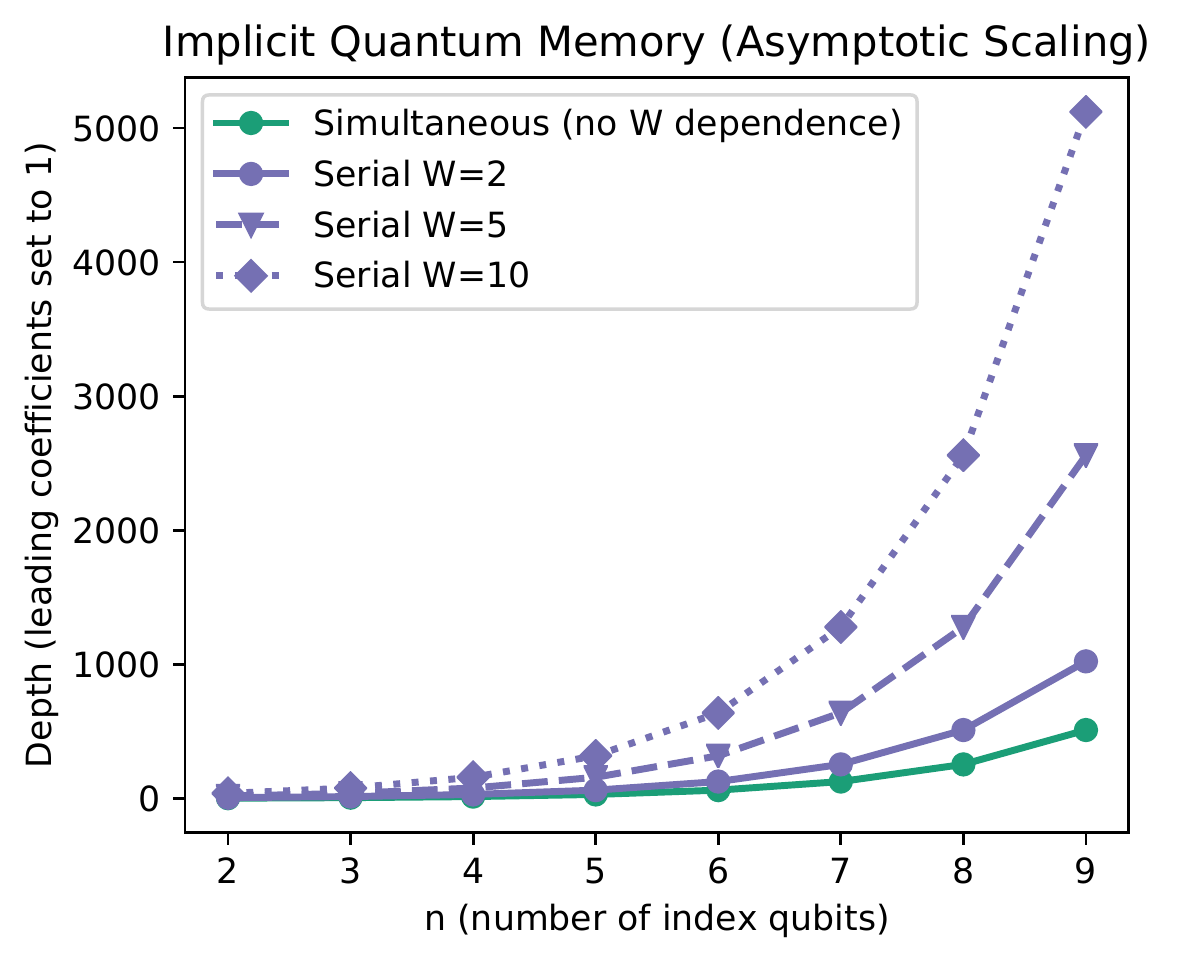}
\end{subfigure}
\hfill
\caption{Depth (lower is better) for SWAP Test, Hadamard Test, and memory architecture benchmarks. We compare circuits compiled with our
Controlled-$U$ circuit synthesis procedure (which uses simultaneous fan-out) versus circuits that serialize the CNOTs. %Specific benchmarks details are documented in \texttt{Benchmarks.ipynb}.
}
\label{fig:depth_results}
\end{figure*}

We evaluated the exact depth reduction for eight applications: SWAP Tests (both traditional and interference circuit), Hadamard Tests (all four applications in Table~\ref{tab:hadamard_test}), and memory architectures (both explicit and implicit). We compiled each benchmark, across a wide range of circuit widths, using both our fan-out based approach (Simultaneous) and the standard serialized approach with no fan-out (Serial). The results are plotted in Figure~\ref{fig:depth_results}.%, and \texttt{Benchmarks.ipynb} has additional details.

We also evaluated the fidelity advantage of simultaneous fan-out for the five most NISQ-friendly benchmarks. For each benchmark type, we found the largest circuit instance with fan-out of at most 8 qubits, matching the largest fan-out we simulated in Figure~\ref{fig:trapped_ion_simulation}. Then, we estimated fidelity with a coarse metric: for each circuit, we assigned each gate a fidelity based on the current hardware ``5\% overrotation, 80 ms laser coherence'' simulation in Figure~\ref{fig:trapped_ion_simulation}. Multiplying together these gate fidelities gives an approximation for the total circuit fidelity (i.e. 1 - infidelity). We also performed this multiplication under the ``1\% overrotation, 80 ms laser coherence'' future scenario with 5x lower overrotation. While these estimates are less accurate than full density matrix simulation, as we performed in Figure~\ref{fig:trapped_ion_simulation}, they are informative from an Amdahl's Law perspective. In particular, single- and two- qubit gates are equally penalized in the Simultaneous and Serial circuits, so the Simultaneous circuits can only perform better when there are large fan-out gates.

\begin{figure}[h]
    \centering
\begin{subfigure}[b]{0.24\textwidth}
\includegraphics[width=\textwidth]{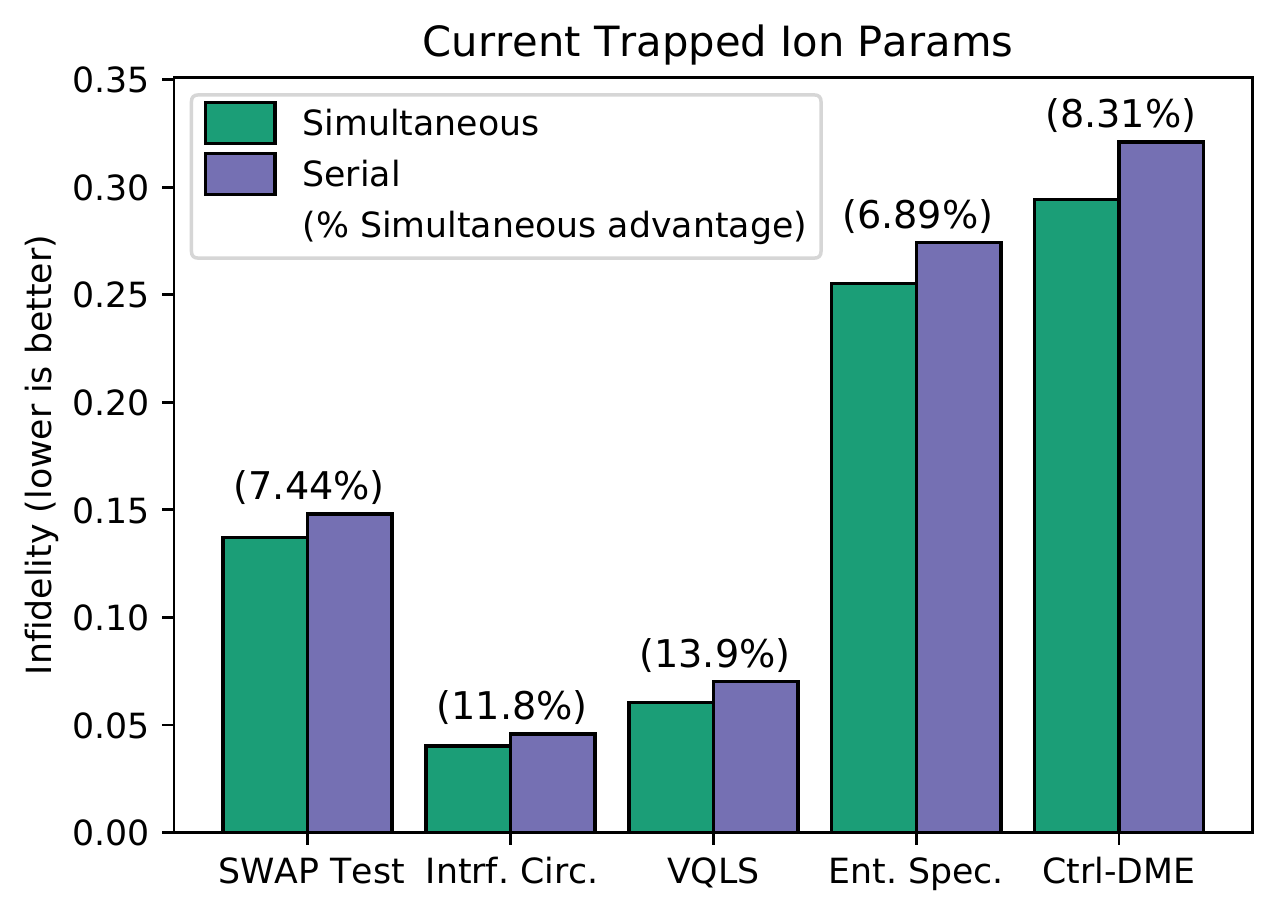}
\end{subfigure}
\hfill
\begin{subfigure}[b]{0.231\textwidth}
\includegraphics[width=\textwidth]{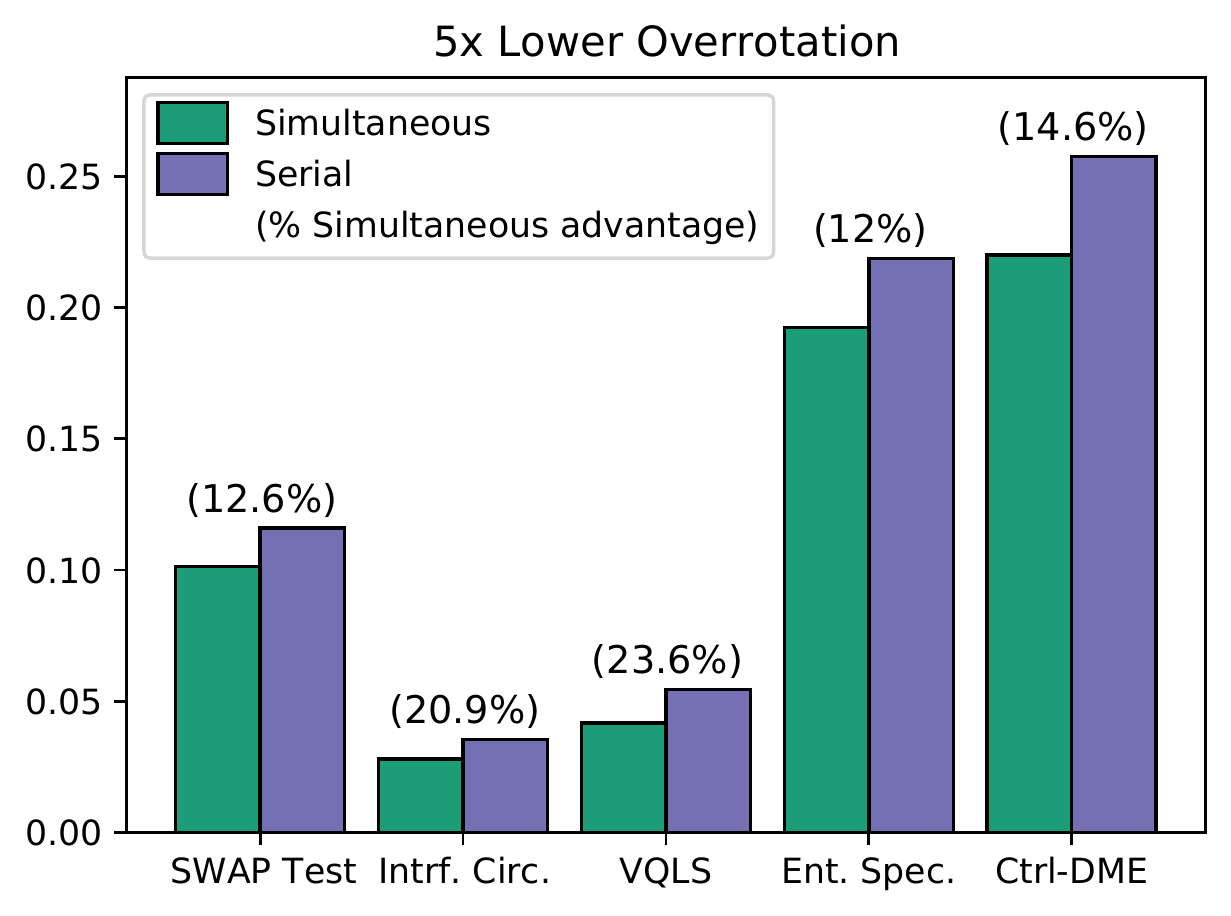}
\end{subfigure}
    \caption{Infidelity estimates for five benchmarks.}
    \label{fig:benchmark_infidelity}
\end{figure}

\subsection{Discussion}

As mentioned in Section~\ref{sec:trapped_ion}, simultaneous fan-out does genuinely give a linear speedup over serialization. Therefore, the depth reductions in Figure~\ref{fig:depth_results} translate directly to faster time-to-solution. This is particularly important since NISQ algorithms need to be variationally executed millions of times \cite{gokhale2019partial}. For four of the eight benchmarked applications, the underlying $U$ has constant depth, so our Simultaneous circuit also has constant depth. For the other four benchmarks, the underlying $U$ has $\Omega(N)$ depth, so both Simultaneous and Serial have increasing depth with $N$. However, Simultaneous' scaling is still lower than Serial's by a linear factor.

The infidelity estimates in Figure~\ref{fig:benchmark_infidelity} have an Amdahl's Law interpretation. The reduction in infidelity from Serial to Simultaneous is greater when the Simultaneous circuit is dominated by fan-out layers. Among our benchmarks, Variational Quantum Linear Solver and Controlled Density Matrix Exponentiation have particularly high fidelity advantages. Our results also demonstrate the sensitivity to the underlying trapped ion hardware's error parameters. For example, VQLS has a 13.9\% Serial$\to$Simultaneous infidelity reduction on current hardware and a 20.9\% reduction on future hardware with 5x lower overrotation.

On current hardware, fidelity is the primary system bottleneck. As such, the fidelity improvement of simultaneous fan-out justifies its use in NISQ machines. 7--24\% reductions in infidelity on 8-qubit circuits are equivalent to months of hardware progress, but our optimization requires no new hardware. As a practical message to hardware providers, we emphasize that exposing global interactions to software will lead to substantial improvements in both fidelity and speed for NISQ applications.
  
\section{Future Work: Superconducting Qubits} \label{sec:superconducting}

Global interaction can be realized for many technologies, but superconducting qubits---which are currently the frontrunner in commercial activity---are a notable exception. To the best of our knowledge, there are no prior implementations of fan-out on superconducting devices. In this section, we demonstrate that superconducting quantum computers can in fact perform simultaneous fan-out. Physical implementation details will be presented in a follow-up paper. %are documented in \texttt{OpenPulse\_Realization.ipynb}.

We first examine the implementation of a CNOT with superconducting qubits. The CNOT is not a natural physical interaction between qubits. Instead, it is performed through a sequence of more primitive physical interactions between qubits. On Google and Rigetti superconducting quantum hardware, CNOT can be realized by a sequence of iSWAP interactions, which are similar to ordinary SWAPs. However, this seems incompatible with simultaneous fan-out, which conceptually requires concurrent reads on the control qubit. By contrast, iSWAP performs both reads and writes on the control qubit since its state is swapped with the target.

An alternative two-qubit interaction called Cross-Resonance \cite{paraoanu2006microwave, rigetti2010fully} is better suited. The Cross-Resonance interaction is used to perform CNOT gates on IBM's devices. It has less restrictive hardware requirements than iSWAP, so Cross-Resonance could be performed on Google and Rigetti hardware as well. Critically, the Cross-Resonance interaction only reads the control qubit, so it does not suffer the immediate barrier to fan-out that iSWAP does.

Although the control qubit \textit{state} is unaffected during Cross-Resonance, the interaction requires (somewhat counterintuitively) driving the control qubit with a microwave pulse. However, by setting this microwave pulse to the frequency of the target qubit, the target qubit will rotate conditioned on the state of the control qubit. This physical interaction easily converts to CNOT through a single-qubit postprocessing gate.

\begin{figure}[h]
    \centering
    \includegraphics[width=0.43\textwidth]{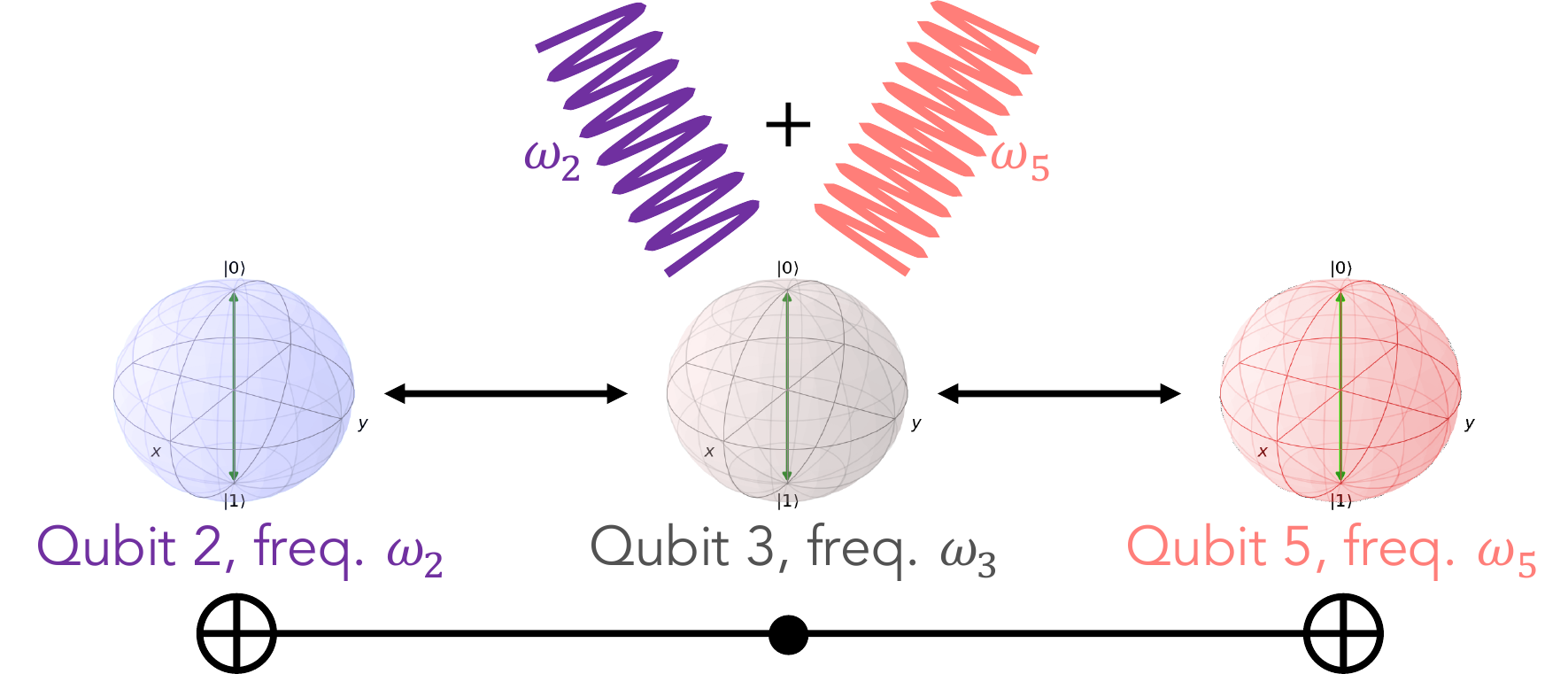}
    \caption{Schematic of fan-out using Cross-Resonance on superconducting qubits. The control qubit (3) is driven with the sum of waves at the targets' frequencies, $\omega_2$ and $\omega_5$.} \label{fig:superconducting_schematic}
\end{figure}

Figure~\ref{fig:superconducting_schematic} illustrates how we can extend this Cross-Resonance interaction to engineer fan-out. In this example, qubit 3 is the control and qubits 2 and 5 are the two targets. To perform the CNOT from 3 to 2 (5), we would drive qubit 3 with microwave at frequency $\omega_2$ ($\omega_5$). However, if we instead drive qubit 3 with the \textit{summation} of two sine waves at frequencies $\omega_2$ and $\omega_5$, then we effectively perform both CNOTs simultaneously. The resulting pulse sequences has a linear speedup over serialization, as desired. Again, this technique only works because Cross-Resonance merely ``reads'' the control qubit, unlike the iSWAP interaction.

We experimentally realized this specific example of fan-out from qubit 3 to qubits 2 and 5 using IBM's Paris quantum computer. We performed our experiment using OpenPulse \cite{mckay2018qiskit, alexander2020qiskit, gokhale2020optimized}, an interface that enables low-level access of quantum computers through Arbitary Waveform Generators. This level of access is required since we need to drive qubit 3 with an unconventional sum-of-waves pulse. We also use sideband modulation, which is needed since the qubit 3 drive is configured to oscillate at $\omega_3$ by default. Moreover, in practice, high fidelity Cross-Resonance interactions require an echo sequence \cite{corcoles2013process} and active cancellation pulses on the target qubits \cite{sheldon2016procedure, magesan2018effective}. Additionally, we had to calibrate a phase offset for the sideband to compensate for accumulated phase on the coaxial cable transitioning from room temperature electronics to the fridge \cite{alexander2020qiskit, krinner2019engineering}. All of these experimental details were handled and will be explained in a follow-up paper.

\begin{figure}[h]
    \centering
    \includegraphics[width=0.48\textwidth]{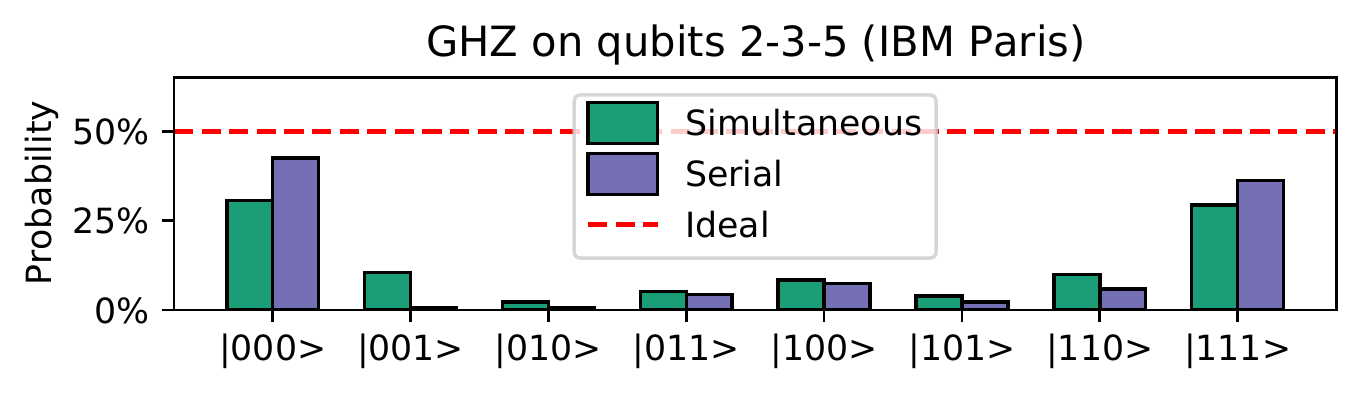}
    \caption{OpenPulse results from $8000 \times 2$ repetitions on IBM Q Paris. The ideal output is 50\% $\ket{000}$ and 50\% $\ket{111}$.}
    \label{fig:openpulse_results}
\end{figure}

Figure~\ref{fig:openpulse_results} shows our experimental results. We attempted to generate the GHZ state, $\frac{\ket{000} + \ket{111}}{\sqrt{2}}$, by first performing a NOT gate on qubit 3 and then fanning out its state to qubits 2 and 5. Ideally, this would result in $\ket{000}$ and $\ket{111}$ each with 50\% probability. With simultaneous fan-out, we achieved 31\% and 29\% respectively. Serialization achieved 42\% and 36\% respectively.

While the GHZ state produced with serial fan-out is better than the one produced with simultaneous fan-out, we emphasize that the simultaneous version ran almost twice as fast. This is encouraging, because superconducting qubits have short coherence lifetimes, so faster operations lead to significant fidelity improvements \cite{gokhale2019partial}. Moreover, when we consider larger width circuits, faster fan-out on a subset of qubits can improve the quality of the other qubits which decohere for less time. Finally, anticipated increases to the sampling rate of Arbitrary Waveform Generators should improve the fidelity of the simultaneous fan-out operation.

Most importantly, our experiment affirms that simultaneous fan-out is possible at all on superconducting quantum hardware. Recent papers have also proposed different techniques that could be used to realize many-body interactions in superconducting systems \cite{rasmussen2020single, chancellor2017circuit, wilkinson2020many, khazali2020fast}, but to our knowledge, our work is the first experimental proof-of-concept. Our work can be viewed a way to engineer crosstalk (unwanted interference between neighboring qubits) for good.

\subsection{Scalability}
An immediate barrier to scaling this simultaneous fan-out procedure to more target qubits is that each control-target pair must be connected in hardware. On superconducting qubit platforms, connectivity is typically sparse. For example, on IBM Q Paris's device topology, the maximum degree is 3, and most qubits are connected to just one or two neighbors. Scaling the connectivity will be a challenge. However, we note that fan-out does not require all-to-all connectivity. Instead, we require a star topology, where a single (control) qubit is connected to every other qubit. Such star topologies have been realized experimentally with 10 qubits connected to a single bus \cite{song201710}. Moreover, star topology is also useful for Hamiltonian simulation circuits \cite{gui2020term}, so there are numerous other quantum subroutines that would also benefit.

A second consideration is that summing waves for each target qubit's frequency (as in Figure~\ref{fig:superconducting_schematic}) will not scale since the maximum amplitude of Arbitrary Waveform Generators is power-constrained. We propose two possible solutions to this. On frequency tunable devices (where $\omega_q$ for each qubit can be controlled), we can simply tune all target qubits to a common frequency during fan-out. Then, the control qubit can be driven at this single common frequency, bypassing the summation of multiple waves. The other solution pertains to fixed-frequency devices. Here, we propose that rectangular-topology qubits could be fabricated with frequencies according to a checkerboard pattern. In such an arrangement, just two colors (frequencies) are needed to ensure no frequency collisions between neighboring qubits. During fan-out, the control qubit can be driven at the sum of just two frequencies, averting the scalability issue.

While these proposed solutions are sound in theory, practical realization will be challenging due to experimental nuances. For example, current qubit fabrication technologies are imprecise and stochastic \cite{brink2018device}, so fabricating qubit frequencies in a checkerboard pattern will be difficult. Thus, more experimental progress will be needed to scale fan-out on superconducting hardware. These hardware-software codesign considerations are valuable in closing the gap from NISQ hardware to practical applications. We propose further work to evaluate simultaneous fan-out with superconducting qubits.

% \section{Preparing Superposition of Inputs}
% Basic idea from Ventura and Martinez \cite{ventura2000quantum} (and maybe others \cite{trugenberger2001probabilistic}) could be improved by:
% \begin{itemize}
%     \item Remove qubits for loading register (may require a N-control Toffoli) and implement in software instead
%     \item For loading step (A/B in Schuld book), implement using single-control multiple targ{\color{blue}}et to make into constant depth
%     \item Maybe...use a bloom filter?
% \end{itemize}

% Can do multi-target via braids or lattice surgery \cite{litinski2017braiding, herr2017lattice, paler2016synthesis}.

\section{Conclusion} \label{sec:conclusion}

At a high level, this work validates the importance of hardware-software codesign. Our core result is driven from the hardware $\rightarrow$ software observation that the exclusive activation structural hazard is not necessary in quantum computing. By exploiting simultaneous fan-out, we are able to synthesized optimized circuit schedules for Controlled-$U$, which is important in NISQ workloads. In the software $\rightarrow$ hardware direction, our results suggest a number of priorities for future hardware development---in particular, the importance of exposing global interactions. Moreover, our demonstration of simultaneous fan-out in superconducting qubits suggests that the star architecture could bring superconducting systems to parity with hardware platforms such as trapped ions.

In current systems, our results affirm a linear speedup from fan-out. In the NISQ era, algorithms will require millions of iterations \cite{gokhale2019asymptotic}, so quantum execution speedups translate to direct reductions in time-to-solution. This opportunity is particularly pronounced on trapped ions, which operate at relatively slow kHz speeds. In addition to the circuit execution speedup, our simulations show 7--24\% infidelity reductions from simultaneous fan-out. This is validated by our trapped ion simulation with a realistic noise model. Our experimental results from superconducting qubits are also promising, though our emphasis is on the mere fact that simultaneous fan-out is possible at all on superconducting qubits.

A number of interesting future directions arise from this work. On the hardware side, we propose experimental realization of our circuits on larger machines, especially in light of recent work noting power law decays for interaction strengths between distant qubits \cite{guo2020implementing}. In addition to superconducting and trapped ion qubits, neutral atom qubits may be promising since global interactions via `Rydberg gates' are natural \cite{muller2009mesoscopic}. On the software side, we propose further investigation of compilation in view of global interactions. \cite{grzesiak2019efficient} suggests that global interactions could in fact give an $O(N^2)$ speedup, whereas we only explore linear speedups in this work. Finding such quadratic speedups could further accelerate the realization of practical quantum computing.

\section*{Acknowledgements}
We are grateful to Ali Javadi-Abhari, Dave Schuster, and Dripto Debroy for helpful suggestions. This work is funded in part by EPiQC, an NSF Expedition in Computing, under grants CCF-1730449/1832377; in part by STAQ under grant NSF Phy-1818914; and in part by DOE grants DE-SC0020289, DE-SC0020331, and DE-SC0019294. We also acknowledge the University of Chicago’s Research Computing Center for their support of this work. Pranav Gokhale is supported by the Department of Defense (DoD) through the National Defense Science \& Engineering Graduate Fellowship (NDSEG) Program.
% Dripto Debroy. Ali Javadi-Abhari and Dave Schuster. Maybe Eric Anschuetz.
% Funding acknowledgements: RCC, EPiQC, STAQ, DOE, NDSEG

%%%%%%% -- PAPER CONTENT ENDS -- %%%%%%%%

%%%%%%%%% -- BIB STYLE AND FILE -- %%%%%%%%
\bibliographystyle{IEEEtranS}
\bibliography{refs}
%%%%%%%%%%%%%%%%%%%%%%%%%%%%%%%%%%%%

\end{document}